\documentclass[showpacs, nobibnotes, amsmath, prd, floatfix,
 reprint, superscriptaddress, apsrev4-1 ]{revtex4-1}

\usepackage{bm}

\usepackage{amssymb}
\usepackage{amsfonts}
\usepackage{graphicx}
\usepackage{rotating}
\usepackage{color}
\usepackage{verbatim}
\usepackage{url}
\usepackage{pifont}
\usepackage{makecell}
\usepackage{placeins}
\usepackage{hyperref}
\usepackage{dcolumn}

\usepackage[displaymath,mathlines]{lineno}
\usepackage{color}

\usepackage{IEEEtrantools}

\begin{document}

\preprint{preprint-number}

\title{Azimuthal transverse single-spin asymmetries of inclusive jets and
charged pions within jets from polarized-proton collisions
at $\sqrt{s} = 500$ GeV}

%\email{James.Drachenberg@lamar.edu}
%\affiliation{Department of Physics, Lamar University}

%Dated: August 17, 2017, 11:29
%Retrieved: August 22, 2017
%http://www.star.bnl.gov/central/collaboration/authors/makeLatex.php

\affiliation{AGH University of Science and Technology, FPACS, Cracow 30-059, Poland}
\affiliation{Argonne National Laboratory, Argonne, Illinois 60439}
\affiliation{Brookhaven National Laboratory, Upton, New York 11973}
\affiliation{University of California, Berkeley, California 94720}
\affiliation{University of California, Davis, California 95616}
\affiliation{University of California, Los Angeles, California 90095}
\affiliation{Central China Normal University, Wuhan, Hubei 430079}
\affiliation{University of Illinois at Chicago, Chicago, Illinois 60607}
\affiliation{Creighton University, Omaha, Nebraska 68178}
\affiliation{Czech Technical University in Prague, FNSPE, Prague, 115 19, Czech Republic}
\affiliation{Nuclear Physics Institute AS CR, 250 68 Prague, Czech Republic}
\affiliation{Frankfurt Institute for Advanced Studies FIAS, Frankfurt 60438, Germany}
\affiliation{Institute of Physics, Bhubaneswar 751005, India}
\affiliation{Indiana University, Bloomington, Indiana 47408}
\affiliation{Alikhanov Institute for Theoretical and Experimental Physics, Moscow 117218, Russia}
\affiliation{University of Jammu, Jammu 180001, India}
\affiliation{Joint Institute for Nuclear Research, Dubna, 141 980, Russia}
\affiliation{Kent State University, Kent, Ohio 44242}
\affiliation{University of Kentucky, Lexington, Kentucky 40506-0055}
\affiliation{Lamar University, Physics Department, Beaumont, Texas 77710}
\affiliation{Institute of Modern Physics, Chinese Academy of Sciences, Lanzhou, Gansu 730000}
\affiliation{Lawrence Berkeley National Laboratory, Berkeley, California 94720}
\affiliation{Lehigh University, Bethlehem, Pennsylvania 18015}
\affiliation{Max-Planck-Institut fur Physik, Munich 80805, Germany}
\affiliation{Michigan State University, East Lansing, Michigan 48824}
\affiliation{National Research Nuclear University MEPhI, Moscow 115409, Russia}
\affiliation{National Institute of Science Education and Research, HBNI, Jatni 752050, India}
\affiliation{National Cheng Kung University, Tainan 70101 }
\affiliation{Ohio State University, Columbus, Ohio 43210}
\affiliation{Institute of Nuclear Physics PAN, Cracow 31-342, Poland}
\affiliation{Panjab University, Chandigarh 160014, India}
\affiliation{Pennsylvania State University, University Park, Pennsylvania 16802}
\affiliation{Institute of High Energy Physics, Protvino 142281, Russia}
\affiliation{Purdue University, West Lafayette, Indiana 47907}
\affiliation{Pusan National University, Pusan 46241, Korea}
\affiliation{Rice University, Houston, Texas 77251}
\affiliation{Rutgers University, Piscataway, New Jersey 08854}
\affiliation{Universidade de Sao Paulo, Sao Paulo, Brazil, 05314-970}
\affiliation{University of Science and Technology of China, Hefei, Anhui 230026}
\affiliation{Shandong University, Jinan, Shandong 250100}
\affiliation{Shanghai Institute of Applied Physics, Chinese Academy of Sciences, Shanghai 201800}
\affiliation{State University of New York, Stony Brook, New York 11794}
\affiliation{Temple University, Philadelphia, Pennsylvania 19122}
\affiliation{Texas A\&M University, College Station, Texas 77843}
\affiliation{University of Texas, Austin, Texas 78712}
\affiliation{University of Houston, Houston, Texas 77204}
\affiliation{Tsinghua University, Beijing 100084}
\affiliation{University of Tsukuba, Tsukuba, Ibaraki, Japan,305-8571}
\affiliation{Southern Connecticut State University, New Haven, Connecticut 06515}
\affiliation{University of California, Riverside, California 92521}
\affiliation{University of Heidelberg}
\affiliation{United States Naval Academy, Annapolis, Maryland 21402}
\affiliation{Valparaiso University, Valparaiso, Indiana 46383}
\affiliation{Variable Energy Cyclotron Centre, Kolkata 700064, India}
\affiliation{Warsaw University of Technology, Warsaw 00-661, Poland}
\affiliation{Wayne State University, Detroit, Michigan 48201}
\affiliation{World Laboratory for Cosmology and Particle Physics (WLCAPP), Cairo 11571, Egypt}
\affiliation{Yale University, New Haven, Connecticut 06520}

\author{L.~Adamczyk}\affiliation{AGH University of Science and Technology, FPACS, Cracow 30-059, Poland}
\author{J.~R.~Adams}\affiliation{Ohio State University, Columbus, Ohio 43210}
\author{J.~K.~Adkins}\affiliation{University of Kentucky, Lexington, Kentucky 40506-0055}
\author{G.~Agakishiev}\affiliation{Joint Institute for Nuclear Research, Dubna, 141 980, Russia}
\author{M.~M.~Aggarwal}\affiliation{Panjab University, Chandigarh 160014, India}
\author{Z.~Ahammed}\affiliation{Variable Energy Cyclotron Centre, Kolkata 700064, India}
\author{N.~N.~Ajitanand}\affiliation{State University of New York, Stony Brook, New York 11794}
\author{I.~Alekseev}\affiliation{Alikhanov Institute for Theoretical and Experimental Physics, Moscow 117218, Russia}\affiliation{National Research Nuclear University MEPhI, Moscow 115409, Russia}
\author{D.~M.~Anderson}\affiliation{Texas A\&M University, College Station, Texas 77843}
\author{R.~Aoyama}\affiliation{University of Tsukuba, Tsukuba, Ibaraki, Japan,305-8571}
% ...

\collaboration{STAR Collaboration}\noaffiliation

\bibliographystyle{apsrev}

\date{\today}

\begin{abstract}
We report the first measurements of transverse single-spin asymmetries
for inclusive jet and $\mathrm{jet}+\pi^{\pm}$ production at midrapidity
from transversely polarized proton-proton collisions at $\sqrt{s} = 500$
GeV. The data were collected in 2011 with the STAR detector sampled
from $23$ pb$^{-1}$ integrated luminosity with an average beam
polarization of $53\%$. Asymmetries are reported for jets with transverse
momenta $6 < p_{T} < 55$ GeV/$c$ and pseudorapidity
$\left|\eta\right| < 1$. Presented are measurements of the inclusive-jet
azimuthal transverse single-spin asymmetry, sensitive to twist-3 initial-state
quark-gluon correlators; the Collins asymmetry, sensitive to quark
transversity coupled to the polarized Collins fragmentation function; and
the first measurement of the ``Collins-like'' asymmetry, sensitive to linearly
polarized gluons. Within the present statistical precision, inclusive-jet and
Collins-like asymmetries are small, with the latter allowing the first
experimental constraints on gluon linear polarization in a polarized proton.
At higher values of jet transverse momenta, we observe the first non-zero
Collins asymmetries in polarized-proton collisions, with a statistical
significance of greater than $5\sigma$. The results span a range of $x$
similar to results from SIDIS but at much higher $Q^{2}$. The Collins results
enable tests of universality and factorization-breaking in the transverse
momentum-dependent formulation of perturbative quantum chromodynamics.
\end{abstract}

\maketitle

\section{Introduction}

The partonic structure of the nucleon at leading-twist in a collinear
picture can be described by three parton distribution functions (PDFs):
the unpolarized parton distribution, $f(x,Q^{2})$; the parton helicity distribution,
$\Delta f(x,Q^{2})$; and the transversity distribution, $h_{1}(x,Q^{2})$
\cite{RalstonSoper,*JaffeJi}. Here, $x$ denotes the parton light-cone momentum fraction,
while $Q^{2}$ denotes the momentum transfer.
The unpolarized PDFs are constrained over a large range of $x$ and $Q^{2}$
by unpolarized lepton-proton experiments at HERA
(e.g.~Ref.~\cite{HERA} and references therein). Helicity distribution
constraints have required observables from the interaction of spin-polarized
probes. Recent global analyses have combined data from polarized
deep-inelastic scattering (DIS), semi-inclusive DIS (SIDIS), and
proton-proton experiments to extract helicity distributions for
quarks, anti-quarks, and gluons
\cite{DSSV08,*DSSV09,*DSSV14,BB,LSS,NNPDF13,*NNPDF14}.
Of the three leading-twist PDFs, transversity, which describes the
transverse polarization of quarks inside a transversely polarized
nucleon, has proven the most difficult to probe. This is due to its
chiral-odd nature that requires transversity to couple to another
chiral-odd quantity in order to be observed.

Advances in understanding transversity have been made through
the study of transverse single-spin asymmetries.
In contrast to expectations based upon collinear perturbative quantum
chromodynamics (pQCD) at leading twist \cite{KPR}, transverse
single-spin asymmetries for inclusive hadron production at forward
pseudorapidity from polarized-proton collisions have long been
observed to be sizable, with asymmetries of a few percent persisting to
transverse momentum ($p_{T}$) values as
high as several GeV$/c$ \cite{ZGSAN,CERNPSAN,
E704-88,*E704,*E704chg,*E704chgpbar,AGSANchg,
STAR_FPD_AN1,*STAR_FPD_AN2,*STAR_FPD2_xSec,
BRAHMSAN,PHENIXMPCAN}. Moreover, the size of the asymmetries
appears to be independent of center-of-mass energy, across
a range of over an order of magnitude. Though the underlying mechanism
driving these effects remains something of a mystery, the
presence of sizable transverse single-spin asymmetries has provided
an opportunity to enhance understanding of pQCD beyond the limits of
collinearity and leading twist.

Non-zero transverse single-spin asymmetries can be generated in pQCD
through the twist-3 and transverse-momentum-dependent (TMD)
formalisms. The twist-3 formalism
\cite{Efremov_twist3_1,QiuSterman_twist3,KanazawaJet,Pitonyak} utilizes
initial-state and final-state multi-parton correlators within the framework of
collinear pQCD. Its application requires a single large-momentum scale parameter and
hence is well-suited for high-$p_{T}$ observables such as high-energy jet
and inclusive hadron production. The TMD formalism \cite{AN_Siv,*AN_Siv2,AN_Col}
utilizes the leading-twist framework of pQCD beyond the collinear
approximation, requiring correlations between spin polarization and
intrinsic transverse momentum. For example, the Sivers mechanism
\cite{AN_Siv,*AN_Siv2} requires a correlation between
the nucleon spin polarization and the intrinsic transverse momentum, $k_{T}$,
of the parton within the nucleon; while the Collins mechanism \cite{AN_Col}
requires a correlation between the polarization of a scattered quark and the
momentum of a hadron fragment transverse to the scattered quark direction.
The Sivers mechanism vanishes in the absence of parton orbital
angular momentum \cite{BrodHwangSchmidt}, and the Collins mechanism is
enabled by nonzero transversity coupled to the polarization-dependent ``Collins''
fragmentation function \cite{AN_Col}.

Unlike twist-3, the TMD approach requires two
momentum scales: a large scale such as $Q$ to enable the use of pQCD, and a soft scale
such as $p_{T}\ll Q$ to enable sensitivity to transverse parton motion.
Moreover, while TMD factorization has been proven for SIDIS as well as for
Drell-Yan and weak-boson production in polarized-proton collisions
\cite{DYFactCollins,TMDFactBoer,TMDFactJi,*TMDFactJi2,
TMDFactCollinsMetz,WANKang}, it is expected that TMD factorization
may generally be broken for hadronic interactions
\cite{TMDFactBreakCollinsQiu,*TMDFactBreakCollins,
TMDFactBreakRogers}. The size of any such factorization breaking
is not known. It has recently been argued that the cross section for
hadrons within jets produced from proton-proton collisions does
factorize and only depends upon universal TMD fragmentation
functions, decoupled from TMD PDFs \cite{TMDFF}. It has also been shown
that as the value of the soft TMD scale becomes larger, the twist-3 and
TMD approaches are mutually consistent
\cite{Twist3TMDConsistent1,*Twist3TMDConsistent2,*Twist3TMDConsistent3}.
Furthermore, the twist-3 correlation functions are related to the
$k_{T}$-integrated TMD distribution and fragmentation functions.
For example, the $k_{T}$-integrated Sivers PDF \cite{AN_Siv,*AN_Siv2}
is related to the Efremov-Teryaev-Qiu-Sterman (ETQS) twist-3 function
\cite{Efremov_twist3_1,QiuSterman_twist3} and similar relations exist
between the Collins and twist-3 fragmentation functions
(e.g.~Refs.~\cite{TMDFactBoer,GambergSivers,Pitonyak}). Consequently,
measurements of twist-3 and TMD observables provide an important
opportunity to test formulations of QCD beyond collinearity and leading twist.

Over the past decade, SIDIS experiments have provided the
first measurements of TMD observables \cite{HERMESColSiv,
*HERMESCol10, HERMESSiv09,COMPASSColSiv05,
*COMPASSColSiv07,*COMPASSColSiv09,*COMPASSColSiv10,
*COMPASSCol12,*COMPASSnew,JLABColSiv}. These, combined
with independent measurements of the Collins fragmentation
function by $e^{+}e^{-}$ experiments \cite{BELLE2006,*BELLE2012,BABAR},
have enabled the first extractions of the transversity PDF
\cite{Transversity07,Transversity09,Transversity13,Transversity15,TransversityKang}.
The kinematic limitations of the current datasets leave the transversity
extractions relatively imprecise for $x\gtrsim0.3$.

\begin{figure}
\begin{center}
\includegraphics[trim = 0mm 5mm 0mm 0mm, clip, width=0.49\textwidth]{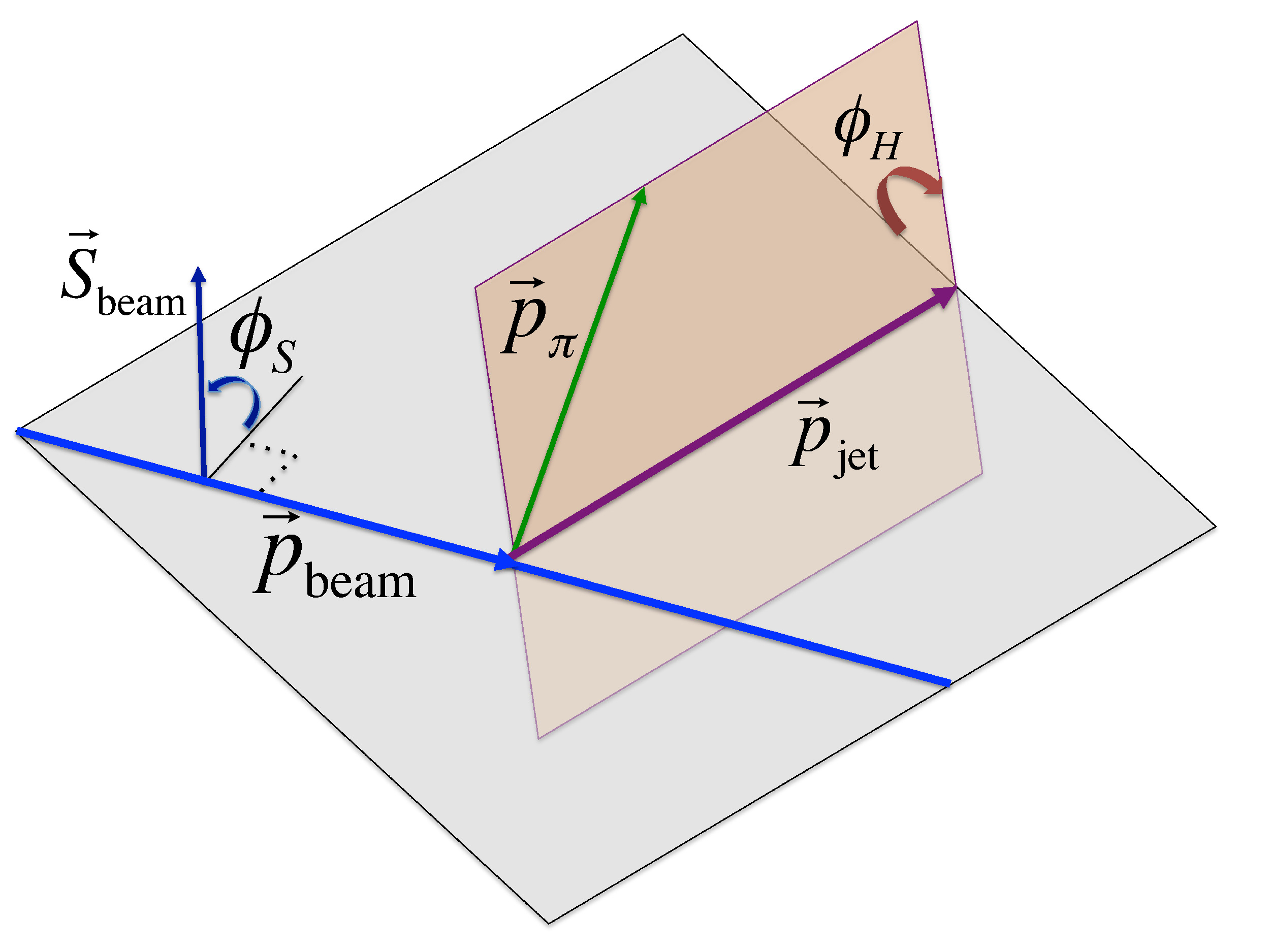}
\end{center}
\caption{\label{fig:angleDef} Azimuthal angle
definitions, following the conventions described in Ref.~\cite{D'Alesio_2011}.
The direction of the beam polarization is denoted by $\vec{S}_{\mathrm{beam}}$,
while the momenta of the polarized beam, jet, and pion are, respectively,
$\vec{p}_{\mathrm{beam}}$, $\vec{p}_{\mathrm{jet}}$, and $\vec{p}_{\pi}$.}
\end{figure}

An incisive way to enhance understanding of
nucleon transverse polarization structure is through the
study of transverse single-spin asymmetries in the production
of jets and pions within jets from polarized-proton collisions
\cite{GPM_2006,YuanCollins,*YuanCollinsLong,D'Alesio_2011}.
The $p_{T}$ of the jet and pion momentum transverse to the
jet axis provide the hard and soft scales, respectively, necessary
for TMD factorization.
By studying different modulations of the transverse single-spin
asymmetry
\begin{eqnarray}
A_{UT}^{\sin\left(\phi\right)}\sin\left(\phi\right) & = & \frac{\sigma^{\uparrow}\left(\phi\right)-\sigma^{\downarrow}\left(\phi\right)}{\sigma^{\uparrow}\left(\phi\right)+\sigma^{\downarrow}\left(\phi\right)} \label{eqn:asymDef},
\end{eqnarray}
one can isolate different physics
mechanisms with sensitivity to various aspects of the nucleon
transverse polarization structure, e.g.~quark transversity and
gluon linear polarization. Measurements with high
energy polarized-proton beams will extend the kinematic reach
in both $x$ and $Q^{2}$ beyond the existing
SIDIS measurements.
The SIDIS cross section scales with the square of the quark
charge, resulting in up quarks being weighted more than down
or strange quarks, a phenomenon often referred to as $u$-quark
dominance. Consequently a large fraction of the observed
$\pi^{-}$ yields arise from the unfavored fragmentation of $u$
quarks. Hadroproduction eliminates $u$-quark dominance,
thereby providing enhanced sensitivity to the minority $d$ quarks.
Furthermore, polarized-proton collisions are directly sensitive to
gluonic subprocesses, enabling the study of the role of gluons in
the transverse polarization structure of the nucleon. Moreover, since
questions remain concerning the magnitude of potential TMD
factorization-breaking in hadronic interactions
\cite{TMDFactBreakCollinsQiu,*TMDFactBreakCollins,
TMDFactBreakRogers,TMDFF}, data from
polarized-proton collisions can provide unique and crucial
experimental insight into these theoretical questions.

Transverse single-spin asymmetries in the production of jets
and pions within jets have a rich structure, as described in
Ref.~\cite{D'Alesio_2011}, the conventions of which we follow
in this article. For pions within jets, the spin-dependent terms in the cross sections 
can be generally expressed \cite{D'Alesio_2011}
\begin{eqnarray}
\lefteqn{d\sigma^{\uparrow}\left(\phi_{S},\phi_{H}\right) - d\sigma^{\downarrow}\left(\phi_{S},\phi_{H}\right)} \nonumber \\
 & \sim & d\Delta\sigma_{0}\sin\left(\phi_{S}\right) \nonumber \\
& + & d\Delta\sigma_{1}^{-}\sin\left(\phi_{S}-\phi_{H}\right) + d\Delta\sigma_{1}^{+}\sin\left(\phi_{S}+\phi_{H}\right) \nonumber \\
& + & d\Delta\sigma_{2}^{-}\sin\left(\phi_{S}-2\phi_{H}\right) + d\Delta\sigma_{2}^{+}\sin\left(\phi_{S}+2\phi_{H}\right) \label{eqn:XsecPol},
\end{eqnarray}
where the $d\Delta\sigma$ terms describe various combinations
of distribution and fragmentation functions. Sinusoidal modulations in particle
production can be measured with respect to two azimuthal
angles: $\phi_{S}$, the azimuthal angle between the proton
transverse spin polarization vector and the jet scattering plane,
and $\phi_{H}$, the azimuthal angle of the pion relative to the
jet scattering plane (Fig.~\ref{fig:angleDef}). The inclusive jet asymmetry, the $\sin(\phi_{S})$
modulation of $A_{UT}$, commonly expressed as $A_{N}$,
is an observable with a single hard scale and therefore driven
by the twist-3 distributions \cite{KanazawaJet}. This observable
is sensitive to the $k_{T}$-integrated Sivers function. The
$\sin(\phi_{S}-\phi_{H})$ modulation of $A_{UT}$
yields sensitivity to transversity coupled to the polarized
Collins fragmentation function.  Through the
$\sin(\phi_{S}-2\phi_{H})$ modulation of $A_{UT}$,
one may gain sensitivity to gluon linear polarization coupled
to the so-called ``Collins-like'' fragmentation function, the
gluon-analog of the Collins fragmentation function. While the
quark-based Collins asymmetry has been measured in
SIDIS, the Collins-like asymmetry has never been measured;
and gluon linear polarization in the polarized proton remains
completely unconstrained. The $\sin(\phi_{S}+\phi_{H})$ and
$\sin(\phi_{S}+2\phi_{H})$ modulations are sensitive to the
TMD transversity distribution and the Boer-Mulders distribution
\cite{BoerMulders} for quarks and gluons, respectively. As
Ref.~\cite{D'Alesio_2011} discusses in detail, these modulations
are not expected to be sizable at the present kinematics, even
under maximized, positivity-bound scenarios. These modulations
are nevertheless measured and found to be consistent with zero.
In principle, the Boer-Mulders distributions for quarks and gluons
may also contribute to the $\sin(\phi_{S}-\phi_{H})$ and
$\sin(\phi_{S}-2\phi_{H})$ modulations. They are, however, again
expected to be negligible under maximized scenarios for the
present kinematics and therefore ignored.

\begin{figure}
\begin{center}
\includegraphics[width=0.47\textwidth]{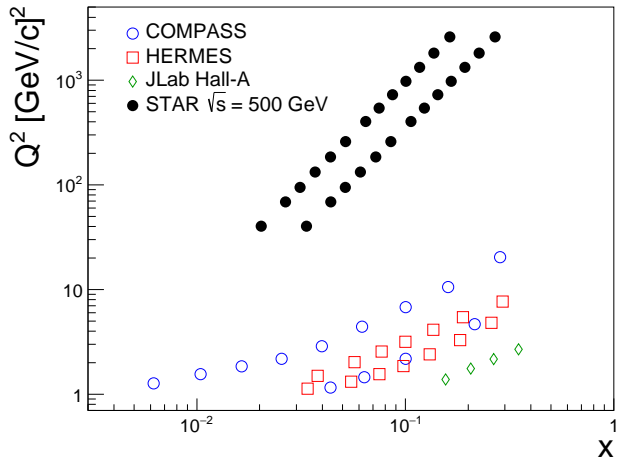}
\end{center}
\caption{\label{fig:x-q2} Range of $x$ and $Q^{2}$ covered
by Collins asymmetry measurements from SIDIS experiments at
HERMES \cite{HERMESColSiv,*HERMESCol10, HERMESSiv09},
COMPASS \cite{COMPASSColSiv05,*COMPASSColSiv07,
*COMPASSColSiv09,*COMPASSColSiv10,*COMPASSCol12,
*COMPASSnew}, and Jefferson Lab Hall-A \cite{JLABColSiv} in
comparison to the $\left(x,Q^{2}\right)$ covered by the present
data as the jet $p_{T}$ and $\eta$ vary.}
\end{figure}

We present the first measurements of the modulations of
$A_{UT}$ for the production of jets and pions within jets
detected in the central pseudorapidity range from polarized-proton
collisions at $\sqrt{s}=500$ GeV. Included are the first measurements
of the Collins-like asymmetry, sensitive to linearly polarized gluons,
and the first observations of the Collins asymmetry in polarized-proton
collisions. For the case of the Collins asymmetry, these measurements
span ranges of $x$ similar to those studied in SIDIS, but at
substantially larger $Q^{2}$, as shown in Fig.~\ref{fig:x-q2}. Additionally,
the inclusive jet asymmetry can lend insight into twist-3 PDFs, such as
the ETQS function, and, thus the Sivers function. In the lower jet-$p_{T}$
range where production is dominated by gluonic subprocesses
(Fig.~\ref{fig:subProcFrac}), the inclusive jet measurements may yield
insight into the quark-gluon and three-gluon twist-3 functions and
potentially the gluon Sivers function that is largely unconstrained
\cite{BoerGluonSivers}. The present data at $\sqrt{s}=500$ GeV provide
improved sensitivity to gluonic subprocess over previous inclusive jet
measurements at $\sqrt{s}=200$ GeV \cite{STAR_jet_AN2}.

\section{Data Analysis}

\subsection{Experiment}
\label{subsec:Experiment}

\begin{figure}
\begin{center}
\includegraphics[width=0.49\textwidth]{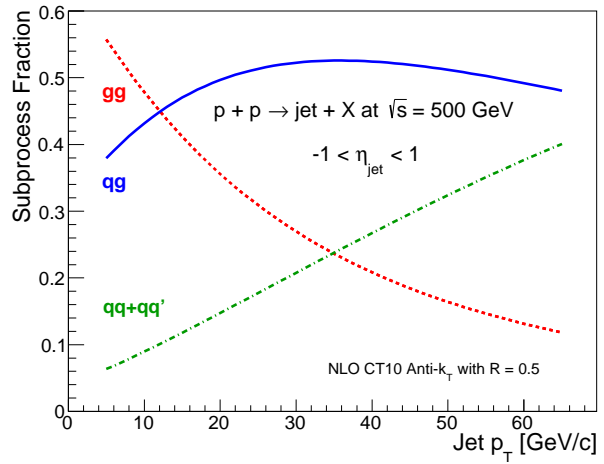}
\end{center}
\caption{\label{fig:subProcFrac} Fraction of next-to-leading order (NLO)
cross section \cite{NLO,CT10} arising from quark-quark (dot-dashed),
gluon-gluon (dotted), or quark-gluon (solid) interactions. The fractions
are shown as a function of jet $p_{T}$ for jets produced within the
pseudorapidity range $-1 < \eta_{\mathrm{jet}} < 1$ from collisions of
an energy $\sqrt{s}=500$ GeV. See Section \ref{subsec:JetReco} for
discussions of the jet definitions and reconstruction.}
\end{figure}

The data were collected during the 2011 run of the Relativistic
Heavy Ion Collider (RHIC) \cite{RHIC_NIM,*RHIC_NIM2} with the
detectors of the Solenoidal Tracker at RHIC (STAR) \cite{STAR_NIM}.
In addition to its role as a heavy-ion collider,
RHIC is the world's only particle accelerator capable of
colliding beams of polarized protons \cite{RHIC-pp_NIM}. The
two independently polarized proton beams are each grouped into 120 
bunches and are loaded with complex patterns of spin direction.
This ensures that systematic effects from asymmetries in spin-dependent
beam luminosities are minimized. RHIC is capable of colliding protons
with polarizations ``longitudinal'' (right-handed or left-handed helicity with respect to the proton momentum)
or ``transverse'' (vertically up or down with respect to the proton momentum).
The beam polarization is measured during the run using Coulomb-nuclear
interference (CNI) polarimetry \cite{2011Pol}. The relative polarization is
measured with proton-carbon polarimeters \cite{pC_Pol} that are
normalized to a hydrogen-jet polarimeter \cite{HJet_Pol} for the absolute
scale. During the 2011 RHIC run, the average beam polarizations were
measured to be $53\%$ and $52\%$ in the clockwise (``blue'') and
counter-clockwise (``yellow'') rotating beams, respectively, as viewed
from above.

The STAR detector is designed to collect data with acceptance
spanning the full azimuth and nearly five units of pseudorapidity,
$-1 \lesssim \eta \lesssim 4$. The present data are generally
collected within the range $\left|\eta\right|<1$. STAR is equipped with detectors
such as the zero-degree calorimeter (ZDC) \cite{STAR_TRG}
for local polarimetry. The ZDC covers the full azimuth for
$\theta<2$ mrad. Detectors such as the vertex position detector
(VPD) \cite{VPD_NIM} are used for minimum-bias triggering. 
The VPD covers approximately half of the solid angle for
$4.2<\eta<5.1$. Among the subsystems utilized in jet
measurements, the time projection chamber (TPC) \cite{TPC_NIM}
provides charged-particle tracking over the full azimuth and
$\left|\eta\right| \lesssim 1.3$. The TPC, supplemented by the
time-of-flight (TOF) system \cite{TOF_NIM} over the range
$\left|\eta\right| \lesssim 0.9$, also provides particle
identification \cite{PID,*PID_highPt}. Surrounding the TPC,
the barrel electromagnetic calorimeter (BEMC) \cite{BEMC_NIM}
provides access to electromagnetic energy over the range
$\left|\eta\right| < 0.98$. This access is extended from $1.09 < \eta < 2.00$
with the endcap electromagnetic calorimeter (EEMC) \cite{EEMC_NIM}.
The BEMC and EEMC are segmented into 4800 and 720 Pb-scintillator
towers, respectively, each corresponding to $\sim20$ radiation lengths and
$\sim1$ nuclear interaction length. The calorimeters also provide
fast input for event triggering. Tracks from the TPC and hits in
the calorimeter towers are also used to reconstruct collision
vertices \cite{STARVtxNote,*STARVtx}. 

\subsection{Event Selection}

All events used in the present analysis were collected from
polarized proton-proton collisions at $\sqrt{s}=500$ GeV. The events are selected from
four different event triggers: a minimum bias trigger requiring a
coincidence in the east and west VPDs (VPDMB) and three different
levels of ``jet-patch'' triggers (JP0, JP1, and JP2). A jet-patch trigger
requires a transverse electromagnetic energy deposit above a certain
threshold--6.4, 9.0, and 13.9 GeV for JP0, JP1, and JP2, respectively--within
a ``patch'' of area $\Delta\eta\times\Delta\phi=1.0\times1.0$ in one of
the calorimeters. The locations of the patches are fixed in space by
the hardware. The jet-patch trigger is discussed in more detail in
Ref.~\cite{STAR_jet_AN2}. In the 2011 run configuration, the jet-patch trigger
geometry consists of thirty patches with overlapping $\eta$ ranges:
$-1<\eta<0$, $-0.6<\eta<0.4$, $0<\eta<1$, $0.4<\eta<1.4$, and $1<\eta<2$.
Due to offline improvements to detector calibrations, in addition to
the hardware trigger, jet-patch triggered events are required to pass
an offline software emulation of the trigger response. A total of 81
million VPDMB events and 23 pb$^{-1}$ of JP2 triggers pass the
basic quality controls for inclusion in the analysis.
The effective luminosities for the pre-scaled JP0 and JP1 triggers fall between these two limits.

Events are required to have a primary collision vertex determined from
the TPC tracking information within a longitudinal position of
$\left|z_{\mathrm{vtx}}\right| < 90$ or $30$ cm of the nominal interaction
point for jet-patch or VPDMB triggers, respectively. For VPDMB events,
this TPC $z_{\mathrm{vtx}}$ is further required to agree within $6$ cm with the
$z_{\mathrm{vtx}}$ determined by the VPD timing information, which removes
pile-up TPC vertices.

\subsection{Jet Reconstruction}
\label{subsec:JetReco}

A comprehensive discussion of jet reconstruction at STAR is presented
in Ref.~\cite{STAR_jet_AN2} (augmented in Ref.~\cite{2009ALL}),
and the present analysis follows a similar approach. Jets are
reconstructed from energetic BEMC and EEMC
towers and from TPC charged-particle tracks passed through the ``anti-$k_{T}$''
jet-finding algorithm \cite{AntiKt} as implemented in the FastJet 3.0.6 C++ package
\cite{FastJet} using a resolution parameter of $R = 0.5$. For inclusion
in the jet reconstruction, TPC tracks must be reconstructed with more than
twelve TPC fit points and with more than $51\%$ of the possible fit points.
These restrictions help to eliminate pile-up tracks, aid momentum resolution,
and eliminate split tracks.
Furthermore, the tracks are required to match within a $p_{T}$-dependent
distance-of-closest-approach to the collision vertex: within 2 cm for
$p_{T} < 0.5$ GeV$/c$, within 1 cm for $p_{T} > 1.5$ GeV$/c$, and
a linearly tapered cut for $0.5 < p_{T} < 1.5$ GeV$/c$. Tracks with
$p_{T} < 0.2$ GeV$/c$ and towers with $E_{T} < 0.2$ GeV are not
included in jet reconstruction. Including the momenta from all
charged particle tracks and energy from all calorimeter towers in jet
reconstruction can lead to overestimations in jet energy due to particles
that leave both a track in the TPC and energy in a calorimeter. To
correct for this overestimation, the track $p_{T}\cdot c$ is subtracted from the
$E_{T}$ of the tower to which it points, not allowing the corrected
$E_{T}$ to be less than 0 \cite{2009ALL}.

\subsection{Jet Selection}

To be selected for analysis, jets are required to have $6 < p_{T} < 55$
GeV$/c$. Furthermore, jets containing tracks with $p_{T,\mathrm{track}} > 30$
GeV$/c$ are excluded to reduce sensitivity to uncertainties in the jet
momentum resolution. To reduce sensitivity to non-collision backgrounds,
such as beam-gas interactions and cosmic rays (observed as neutral
energy deposits in the calorimeters), jets are required to derive at least
$6\%$ of their energy from charged particles with no less than $0.5$
GeV$/c$ in total charged-particle $p_{T}$. Owing to the asymmetric
electromagnetic calorimeter coverage, jets are required to fall
within an asymmetric window of pseudorapidity,
$-0.7 < \eta_{\mathrm{detector}} < 0.9$, with respect to the
center of the detector. Additionally, the jets are required to fall within
a window of $\left|\eta\right|<1.0$ from the reconstructed collision
vertex. Jets from jet-patch triggered events are required to satisfy a
geometric match to a hardware jet patch with energy deposition
above the nominal triggering threshold. For the present analysis,
this requirement is
$\left|\eta_{\mathrm{jet}}-\eta_{\mathrm{patch}}\right| < 0.6$ and
$\left|\phi_{\mathrm{jet}}-\phi_{\mathrm{patch}}\right| < 0.6$.
Furthermore, jets are required to have a minimum $p_{T}$ of 7.1,
9.9, or 16.3 GeV$/c$ for JP0, JP1, and JP2 events, respectively.

\subsection{Pion Selection}
\label{subsec:PionSelect}

Charged pions are chosen for the Collins and Collins-like analyses
from jets passing the aforementioned selection criteria by requiring
additional restrictions beyond those applied to all particles for inclusion
in jet reconstruction. These pions are required to have $0.1 < z < 0.8$,
where $z$ is the ratio of pion momentum to jet momentum (Note that
this definition for $z$ is referred to as $z_h$ in Ref.~\cite{TMDFF}).
A significant portion of pions with $z<0.1$ in low $p_T$ jets arise from
underlying event backgrounds, while $z=0.8$ marks the limit for
sufficient statistics in this dataset. For simplicity, a single range
of $z$ is selected independent of jet $p_T$.
To ensure robust reconstruction of the relevant azimuthal angles, pions
are required to fall outside a radius of $\Delta R > 0.04$, where
\begin{equation}
\Delta R = \sqrt{\left(\eta_{\mathrm{jet}}-\eta_{\pi}\right)^{2}+\left(\phi_{\mathrm{jet}}-\phi_{\pi}\right)^{2}} \label{eqn:DeltaR}
\end{equation}
relative to the jet axis. This requirement is discussed further in
Section \ref{subsec:AziRes}. To limit contamination to the pion
sample from kaons, protons, and electrons, pions are only
selected for analysis if the observed $dE/dx$ is consistent with
the expected value for pions. A parameter $n_{\sigma}(\pi)$ is
defined as
\begin{eqnarray}
n_{\sigma}\left(\pi\right) & = & \frac{1}{\sigma_{\mathrm{exp}}}\ln\left(\frac{dE/dx_{\mathrm{obs}}}{dE/dx_{\pi\; \mathrm{calc}}}\right) \mbox{,}
\end{eqnarray}
where $dE/dx_{\mathrm{obs}}$ is the observed value for
the event, $dE/dx_{\pi\; \mathrm{calc}}$ is the expected
mean $dE/dx$ for pions of the given momentum, and
$\sigma_{\mathrm{exp}}$ is the $dE/dx$ resolution of the
TPC \cite{PID,PID_highPt}. To ensure reliable particle
identification, pions are further required to contain at least
six TPC fit points with valid $dE/dx$ information. The $dE/dx$
values for kaons and protons overlap those of pions at momenta
of 1.1 GeV$/c$ and 1.7 GeV$/c$, respectively, in the STAR TPC.
Thus, the $n_{\sigma}\left(\pi\right)$ selection window is varied
depending on the reconstructed particle kinematics to optimize
the sample purity. The background correction procedure
and sample purity are described in Section \ref{subsec:NonPionBG}.

\subsection{Spin Asymmetry Analysis}
\label{subsec:SpinAsymAnalysis}

The transverse single-spin asymmetry, defined in Eq.~\ref{eqn:asymDef},
is the amplitude of the sinusoidal $\phi$-dependence of the
spin-dependent cross section, where $\phi$ represents any of the
relevant azimuthal angles, e.g.~for the inclusive-jet, Collins, or Collins-like
effects. For extraction, the present analysis utilizes the so-called
``cross-ratio'' method \cite{CrossRatio} for which both acceptance and
luminosity effects cancel to first order. The cross-ratios, $\epsilon$, for a given bin
of $\phi$ can be formed by combining yields from azimuthally opposite detector halves
($\alpha$ vs.~$\beta$) when the spin orientations are flipped
\begin{eqnarray}
\epsilon=\frac{\sqrt{N_{1,\alpha}^{\uparrow}N_{1,\beta}^{\downarrow}}-\sqrt{N_{1,\alpha}^{\downarrow}N_{1,\beta}^{\uparrow}}}{\sqrt{N_{2,\alpha}^{\uparrow}N_{2,\beta}^{\downarrow}}+\sqrt{N_{2,\alpha}^{\downarrow}N_{2,\beta}^{\uparrow}}} \label{eqn:crossRatioWeight} \mbox{,}
\end{eqnarray}
where $N_{1}$ is the particle yield for a given spin state in each detector
half, weighted by the beam polarization for the event. $N_{2}$ is the same,
though weighted by the square of the polarization. For each RHIC fill, an initial
polarization ($P_{0}$), initial time ($t_{0}$), and decay slope ($dP/dt$) for
each beam are provided by the CNI polarimeter \cite{2011Pol}. The
event-by-event polarizations are calculated as
$P_{0} + \left(dP/dt\right)\times\Delta t$, where $\Delta t$ is the
difference between $t_{0}$ and the event timestamp. An overall
systematic uncertainty of $\sigma\left(P\right)/P = 3.5\%$ is assigned
to both blue and yellow beams \cite{2011Pol} and is considered an uncertainty
in the vertical scale of the asymmetries.

For the present analysis, events are designated $\alpha$ or $\beta$
for jets in the ``top'' ($0 < \phi_{\mathrm{jet}} < \pi$) or ``bottom''
($-\pi < \phi_{\mathrm{jet}} < 0$) half of the TPC, respectively. The
cross ratios are formed as functions of the azimuthal angles, and
the raw asymmetries are extracted using fits of the form
\begin{eqnarray}
\epsilon\left(\phi\right) & = & a+b\sin\left(\phi\right) \mbox{.}
\end{eqnarray}
The parameter $b$ is the raw asymmetry, while the $a$ term
is expected to be zero. It is left in the fit as a
cross-check for hidden systematic effects and found to be consistent
with zero in all cases. Further statistical checks are the examination
of the fit $\chi^{2}$ and asymmetry residual distributions. The fits
are initially performed separately for each RHIC fill, event trigger,
and RHIC beam. $\chi^{2}$ distributions for the asymmetry extractions
are evaluated by fitting with a $\chi^{2}$ function allowing the parameter
for the number of degrees of freedom, $\nu$, to float. In all cases, the fits
to the $\chi^{2}$ distributions return values of $\nu$ consistent with the expectation from
the relevant number of azimuthal angle bins and fit constraints. The weighted
averages of the asymmetries are formed and the residual distributions
are examined for signs of outliers or underestimated uncertainties.
In each case, the residuals follow the expected form of a Gaussian
with unit width centered at zero, suggesting the individual fits scatter
statistically about the mean with well estimated uncertainties. For the
final extractions, yields from all of the fills and the blue and yellow beams
are combined to maximize the statistical precision. To allow for
trigger-dependent corrections, the asymmetries are extracted
trigger-by-trigger; and the corrected asymmetries are combined in
a weighted average over the event trigger classes.

\begin{figure}
\begin{center}
\includegraphics[width=0.49\textwidth]{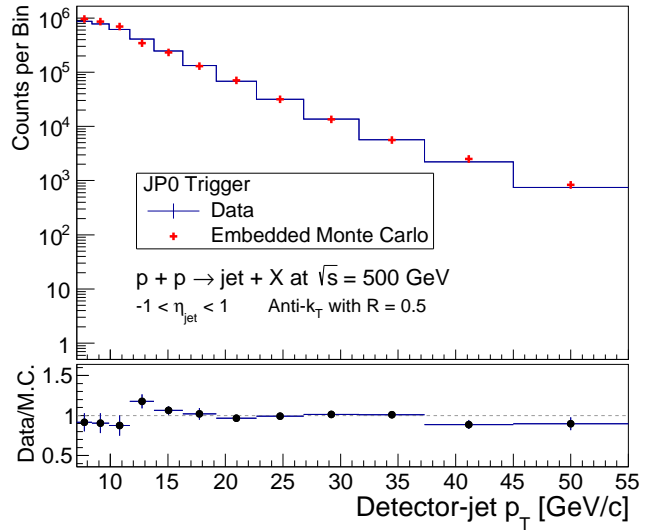}
\end{center}
\caption{\label{fig:DataMCpT} Comparison of data to embedded
Monte Carlo simulations for data and simulated events that satisfy the JP0 trigger. The
embedding Monte Carlo distribution is normalized to match the number of data
counts across the range $16.3 < p_{T} < 45.0$ GeV$/c$.}
\end{figure}

\begin{figure*}
\begin{center}
\includegraphics[width=0.98\textwidth]{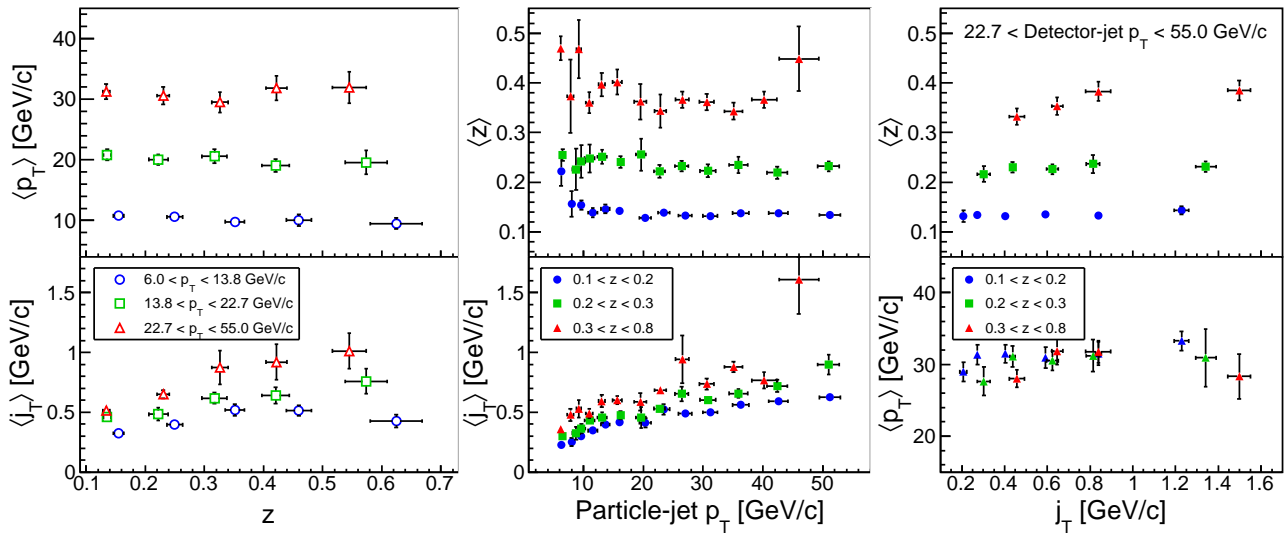}
\end{center}
\caption{\label{fig:MeanKine} Mean corrected pion $z$, jet $p_{T}$,
and pion $j_{T}$ as a function of each kinematic variable for pions within jets. Reconstructed
kinematic variables are corrected to their underlying values at the PYTHIA particle-jet
level. Uncertainties are the quadrature combinations of statistical and systematic contributions.}
\end{figure*}

\section{Systematic Corrections and Uncertainties}

\subsection{Simulations}

To evaluate systematic effects such as shifts in the jet or particle
kinematics or trigger and reconstruction bias, QCD events are
produced with the PYTHIA Monte Carlo generator \cite{Pythia}, fed
through a simulated detector response, and their resulting ADC counts
embedded into real events, triggered randomly (zero-bias) during a
nominal crossing of proton bunches. This ``embedding''
procedure enables accounting of such effects as out-of-time pile-up
that are otherwise difficult to simulate. The Monte Carlo events are generated
with PYTHIA 6.426 \cite{Pythia} using the Perugia 0 tune \cite{Pythia2}. In
order to achieve an optimal reproduction of the calculated next-to-leading order (NLO) inclusive
jet transverse momentum spectrum, the intrinsic partonic $k_{T}$ parameter
is set to 1 GeV$/c$. The detector response is
emulated with the GSTAR package based upon GEANT 3.21/08
\cite{GEANT0,*GEANT1}. The individual data-taking runs into which the
PYTHIA events are embedded are chosen so as to sample the full range
of luminosities encountered during the 2011 RHIC run. The runs were
selected to cover a broad span of time, so that changes in detector
hardware states would be well represented in the simulation sample.

The same analysis software and jet-finding algorithms applied to the
data are utilized to reconstruct jets from the embedded Monte Carlo.
The jet-finding is applied on three levels: At the ``detector-jet'' level, jets
are formed from TPC tracks and calorimeter towers after full simulation
of the detector response. Detector jets allow direct contact between the
data and simulation and exhibit excellent agreement (e.g.~Fig.~\ref{fig:DataMCpT}).
PYTHIA ``particle-jets'' are constructed by applying the anti-$k_{T}$ algorithm
with $R=0.5$ to all stable, hadronized, final-state particles from the simulated
collision, prior to detector simulation. PYTHIA ``parton-jets'' are constructed by
applying the anti-$k_{T}$ algorithm with $R=0.5$ to hard-scattered partons
including initial-state and final-state radiation but excluding beam remnant
and underlying event contributions.

\subsection{Corrections to Jet and Pion Kinematics}
\label{subsec:KineCorr}

Detector jet and pion kinematics are corrected to the PYTHIA
particle-jet level. Detector jets are associated in $\left(\eta,\phi\right)$
with both a particle-jet and a parton-jet in the event with the association
defined as $\sqrt{\Delta\eta^{2} +\Delta\phi^{2}} < 0.5$.
Similarly, pions from associated detector-jets are further associated
with particles at the PYTHIA particle-jet level. Asymmetries are
presented at the mean corrected kinematic values for each detector-jet
level bin, determined by the kinematics of the associated PYTHIA jet. A summary of the corrected values of pion $z$, jet $p_{T}$,
and pion $j_{T}$ (pion momentum transverse to the jet axis) is
presented in Fig.~\ref{fig:MeanKine}.

Uncertainties in the corrections for jet $p_{T}$ and $z$ are the quadrature
combinations of statistical and systematic contributions, including
those due to Monte Carlo statistics, calorimeter gains, calorimeter
response to charged particles, TPC tracking efficiency, track momentum
resolution, and trigger emulation. Uncertainties in the corrections for
pion $j_{T}$ exclude those of the jet energy scale.

\subsection{Parton-jet Associations}
\label{subsec:Match}

At lower values of jet $p_{T}$, a non-negligible probability exists that a
reconstructed jet arose not from the hard scattering process but
either from the underlying event or from pile-up backgrounds. These
probabilities are estimated by analyzing jets reconstructed from the
embedded Monte Carlo events. First, an association is required
between the reconstructed and generated event vertices. Association
probabilities range from $\sim90-95\%$ at the lowest values of
jet $p_{T}$ to unity for $p_{T} > 10$ GeV$/c$. An association is
further required between the reconstructed detector-jet and a PYTHIA
particle-jet. These association probabilities range from
$\sim95\%$ at the lowest $p_{T}$ values, rising to unity by $\sim10$
GeV$/c$. Finally, an association is required between the reconstructed
jet and a PYTHIA parton-jet. For jets with $p_{T} > 10$
GeV$/c$, parton-jet association probabilities are near unity. Below
$10$ GeV$/c$, these association probabilities decrease steadily to
$\sim75\%$ at $6$ GeV$/c$. For purposes of the association, the
matched parton or particle jets are required to have at least $1.5$
GeV$/c$ of transverse momenta.

For the asymmetries studied in the present analysis, the pertinent
physics mechanisms are largely parton-level effects. For asymmetries
of pions within jets, hadronization also plays a critical role. It is unclear
what effects may be induced by the presence of jets from the underlying
event and beam remnant; thus, the present analysis does not correct for the
association probabilities. Instead, a systematic uncertainty
is applied which considers the full difference between the measured
asymmetry and the value if a dilution correction had been applied.

\begin{table}
\begin{center}
\resizebox{0.45\textwidth}{!}{
\begin{tabular}{lcc}
\hline\hline
$p_{T}$ Range [GeV$/c$] & Quark Bias & Gluon Bias\\ \hline
$6.0-13.8$ & $+17\%$ & $-10\%$ \\ 
$13.8-22.7$ & $+8\%$ & $-6\%$ \\ 
$22.7-55.0$ & $+3\%$ & $-3\%$ \\ \hline\hline
\end{tabular}
}
\end{center}
\caption{\label{tab:TrigBias}Trigger bias systematics for quark and
gluonic subprocesses. The values reflect the relative difference
in the fraction of quark or gluon jets between triggered and unbiased
Monte Carlo events. The quark biases apply to the Collins asymmetries
while the gluon biases apply to the inclusive-jet and Collins-like
asymmetries. The systematics are correlated across the full range of jet $p_{T}$.}
\end{table}

\subsection{Trigger Bias}

For a given jet $p_{T}$, in particular at low $p_{T}$, the fixed size of
the jet patches leads to a higher trigger efficiency for quark jets than
less-collimated gluon jets (see the discussions in
Refs.~\cite{STAR_jet_AN2,2009ALL}). For the inclusive jet and
Collins-like asymmetries, each of which are largely driven by gluonic
effects, the trigger bias will suppress the asymmetries. For
the Collins asymmetry, the bias should serve to enhance the effect.

The trigger bias effects are estimated with Monte Carlo simulations by
comparing fractions of quark and gluon jets in triggered and unbiased
event samples. In the embedded Monte Carlo, detector jets are matched
to hard-scattered partons and sorted into quark and gluon jets. The
unbiased jet sample, free of trigger, pile-up and reconstruction effects,
is constructed from all particle jets that fall within the nominal $p_{T}$
and $\eta$ kinematic ranges. The trigger-bias estimates are summarized
in Table \ref{tab:TrigBias}. The associated systematic uncertainties are
calculated by scaling the measured asymmetry by the bias fractions from
Table \ref{tab:TrigBias} and are correlated across the full range of jet $p_{T}$.

\begin{figure}
\begin{center}
\includegraphics[width=0.49\textwidth]{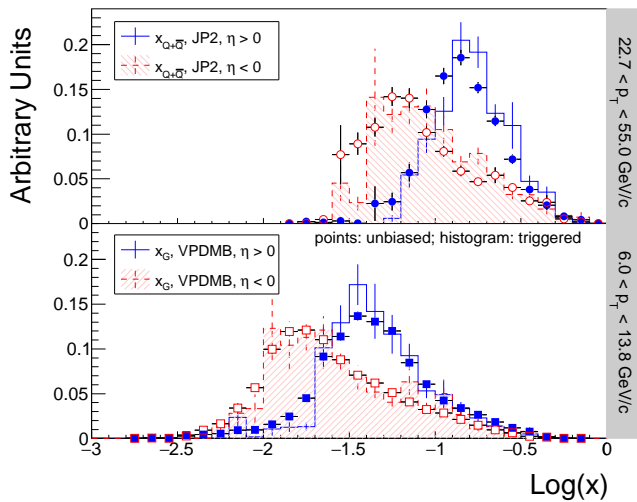}
\end{center}
\caption{\label{fig:TrigBiasX}
Underlying parton $x$ distribution for all Monte Carlo simulated jets
(points) within the specified range of $p_{T}$ compared to those
passing the indicated event trigger condition (histogram). Uncertainties
are from Monte Carlo statistics. The upper
panel shows quark $x$ values for high-$p_{T}$ jets, comparing JP2
to unbiased events, while the bottom panel presents gluon $x$ values
for low-$p_{T}$ jets, comparing VPDMB to unbiased events. For JP2,
deviations can arise both from reconstruction effects and from bias in
the event trigger. For VPDMB events, deviations from the unbiased
sample arise from reconstruction and pile-up effects.}
\end{figure}

In addition to the bias in partonic subprocess fraction, the active
range of $x$ will be somewhat distorted by the effects of
reconstruction and trigger bias. To estimate these effects, the embedding
simulations are analyzed, again, using the jets matched to
the hard-scattered partons. The hard scattered parton is further
required to match the flavor of one of the two partons that
initiated the hard-scattering process. Events are separated based
on trigger, jet $p_{T}$, parton species, and $\eta$ calculated
relative to the incident parton direction. The effect of the bias is
demonstrated in Fig.~\ref{fig:TrigBiasX}, where the triggered and
reconstructed embedded jets are compared to those with trigger,
reconstruction, and detector effects removed.
Results for two cases, $x_{G}$ for low-$p_{T}$ VPDMB and
$x_{Q}$ for high-$p_{T}$ JP2, are shown in Fig.~\ref{fig:TrigBiasX}.
The VPDMB distribution agrees quite well with the
unbiased distribution, while the JP2 distribution is shifted slightly
to higher $x$, relative to the unbiased distribution. These deviations
can arise from biases in both reconstruction and event triggering. Uncertainties
reflected in Fig.~\ref{fig:TrigBiasX} are from Monte Carlo statistics.

\subsection{Azimuthal Resolutions}
\label{subsec:AziRes}

Since the asymmetries of interest are extracted from the azimuthal
dependence of the spin-dependent cross sections, finite azimuthal
resolution leads to a systematic dilution of the true asymmetries.
There are two main contributors to the finite resolution: inefficiencies
in TPC tracks and calorimeter towers, which distort the reconstruction
of the jet thrust axis, and errors in the reconstructed
positions of the tracks and towers. An additional resolution effect,
finite azimuthal bin width, is discussed in Section \ref{subsec:Acceptance}.
The size of the asymmetry smearing is estimated by evaluating the
distribution of event-by-event deviations of the reconstructed azimuthal
angles from their true value in the embedded simulation. The
resulting distribution is convolved with a sinusoid, and the ratio
of the resulting amplitude to the original is taken as the size of the
dilution due to finite azimuthal resolution. This dilution is used to
correct the measured asymmetries. Uncertainties for the resolution
correction include both statistical contributions (from finite Monte
Carlo statistics) and systematic contributions (e.g.~from the
accuracy of the Monte Carlo simulations).

The RMS of the distribution of event-by-event jet-axis deviations
is $\sim0.4$ radians at the lowest jet $p_{T}$ values and decreases
to $\sim0.15$ radians at high jet $p_{T}$. For inclusive-jet
observables, this translates to finite azimuthal resolution effects of
less than $10\%$. In low-$p_{T}$ Collins and Collins-like asymmetries,
smearing effects can become more significant. Enhanced
smearing of the jet axis at low $p_{T}$ due to less-collimated jets is
compounded by the azimuthal resolution of the pion orientation. This
carries with it a potentially strong dependence upon the proximity
of the pion to the jet axis, denoted here as $\Delta R$ (Eq.~\ref{eqn:DeltaR}).
To ensure robust reconstruction of the relevant angles, a minimum
cut on $\Delta R$ is imposed for the Collins and Collins-like
asymmetries. The cut is optimized by balancing the degradation
in azimuthal resolution with the loss in statistics. Furthermore, $z$ and
$j_{T}$ become tightly correlated for more restrictive $\Delta R$
cuts. This is of crucial importance to the Collins asymmetry measurement
which may have strong dependence upon both $z$ and $j_{T}$. Specifically,
the minimum $j_{T}$ may be approximated
\begin{eqnarray}
j_{T,\, \mathrm{min}} & \approx & z\times\Delta R_{\mathrm{min}}\times p_{T,\, \mathrm{jet}}\mbox{,}
\label{eqn:RminJt}
\end{eqnarray}
where $\Delta R_{\mathrm{min}}$ is the minimum $\Delta R$,
and $p_{T,\, \mathrm{jet}}$ is the $p_{T}$
of the jet. Thus, the $\Delta R$ cut introduces a more stringent
limit on $j_{T}$ at higher values of jet $p_{T}$, where the present
data are expected to have the most sensitivity to the Collins effect.
For the analysis, $\Delta R_{\mathrm{min}} = 0.04$ is chosen. The
resulting RMS of the distribution of event-by-event $\phi_{H}$
deviations is $\sim0.78$ radians at the lowest jet $p_{T}$ values and
decreases to $\sim0.40$ radians at high jet $p_{T}$. 

\subsection{Non-uniform Acceptance Effects}
\label{subsec:Acceptance}

The jet and pion yields for the various spin states depend
upon all of the possible azimuthal modulations
\cite{D'Alesio_2011,KanazawaJet}. Twist-3 distributions (and
indirectly the Sivers function) can give rise to modulations
of the form $\sin(\phi_{S})$. Transversity coupled to
the polarized Collins fragmentation function can give
rise to modulations of the form $\sin(\phi_{S}-\phi_{H})$.
Gluon linear polarization coupled to the polarized
Collins-like fragmentation function can give rise to
modulations of the form $\sin(\phi_{S}-2\phi_{H})$. In
principle, modulations of the form $\sin(\phi_{S}+\phi_{H})$
and $\sin(\phi_{S}+2\phi_{H})$ are also possible;
but with the kinematic range of the present analysis they are
expected to be negligible, even under maximized,
positivity-bound scenarios \cite{D'Alesio_2011}. Therefore,
the corrections and systematic contributions are neglected
in the present analysis for the latter two modulations.
Nevertheless, their raw modulations have been examined
and are found to be consistent with zero, and any corresponding
systematic effect could constitute no more than 10-20\% of the
total systematic uncertainty. In the limit
of uniform acceptance, the cross-ratios described
in Section \ref{subsec:SpinAsymAnalysis} will isolate the
desired observables, decoupled from the competing
asymmetry modulations. The STAR detector has excellent
uniformity.  Nonetheless, small distortions in the detector
response can couple to non-zero competing physics
asymmetries to distort the extraction of the modulation
of interest.  These ``leak-through'' effects can manifest
as a sinusoidal dependence in the angular distribution
of interest or as higher-order Fourier components, depending
on the combination of observables and level of
non-uniform instrumental acceptance.

\begin{figure*}
\begin{center}
\includegraphics[width=0.98\textwidth]{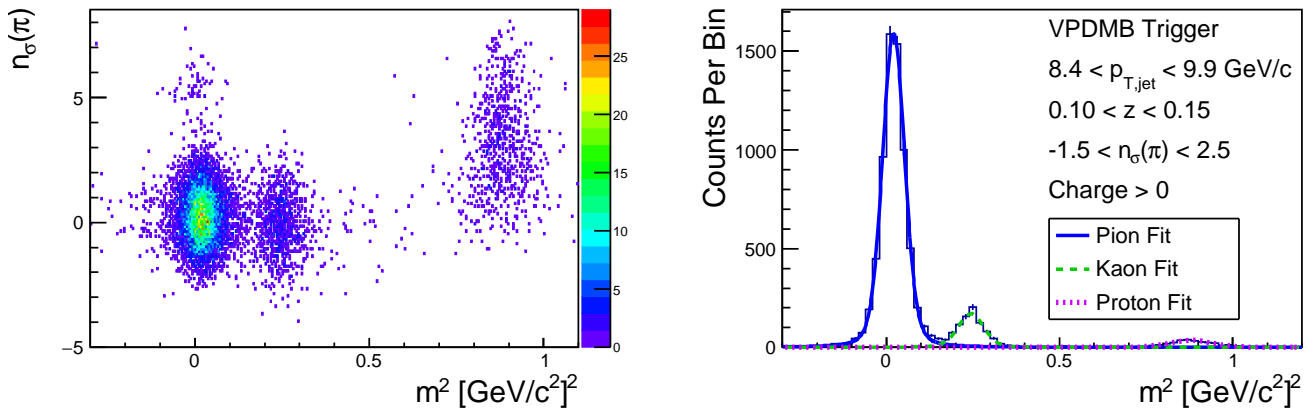}
\end{center}
\caption{\label{fig:TOF} $n_{\sigma}\left(\pi\right)$
vs.~$m^{2}$ (left) and a multi-Voigt-profile fit to $m^{2}$ (right)
for VPDMB-triggered events with positively charged particles
carrying momentum fractions of $0.10 < z < 0.15$ from jets with
$8.4 < p_{T} < 9.9$ GeV$/c$. For the example fit shown, the charged
particles are required to satisfy $-1.5 < n_{\sigma}\left(\pi\right) < 2.5$.
The individual contributions to the overall fit from pions, kaons, and
protons are shown.}
\end{figure*}

The shape of the leak-through distortion is extracted
from the data itself, using a technique motivated by
the conventional mixed-event procedure that is
frequently used to measure acceptance effects. As
an example, consider the distortion to the Collins
modulation, $\phi_{S}-\phi_{H}$, due to the
inclusive-jet asymmetry. Yields are binned for each
event, in the same manner as with the asymmetry
analysis. Events are given weights of
\begin{eqnarray}
w_{0} & = & 1+A_{\mathrm{in}}\sin\left(\phi_{S}\right) \label{eqn:ColSivPos} \\
w_{1} & = & 1-A_{\mathrm{in}}\sin\left(\phi_{S}\right) \label{eqn:ColSivNeg}\mbox{,}
\end{eqnarray}
where $A_{\mathrm{in}}$ is an input asymmetry. Each event
is binned twice, once for each proton spin state. The histogram
corresponding to the actual spin state for the event is filled
using weights of $w_{0}$, while the histogram for the opposite
spin state is filled using weights of $w_{1}$. In this manner, an
unpolarized sample is constructed (in the limit of vanishing
luminosity asymmetries) removing the possibility for
complications from actual physics coupling to the input
asymmetries. A similar procedure is applied for each of the
desired physics effects and their possible contaminations.
The cross ratios for the relevant asymmetries are calculated
using the weighted-event histograms and are fit with
a function of the form
\begin{eqnarray}
\epsilon\left(\phi\right) & = & p_{0}+p_{1}\sin\left(\phi\right)
\end{eqnarray}
(Section \ref{subsec:SpinAsymAnalysis}). The relevant
distortions to the asymmetries of interest materialize in the $p_{1}$ parameter of the fit. A
conservative systematic uncertainty due to finite
acceptance effects is estimated as
\begin{eqnarray}
\sigma_{\mathrm{leak}} & = & \mathrm{Max}\left(\left|A_{\mathrm{leak}}\right|,\sigma_{A_{\mathrm{leak}}}\right)\times\frac{\mathrm{Max}\left(\left|p_{1}\right|,\sigma_{p_{1}}\right)}{A_{\mathrm{in}}} \mbox{,}
\end{eqnarray}
where $A_{\mathrm{leak}}$ is the measured
asymmetry for the ``competing'' effect and
$A_{\mathrm{in}}$ is the asymmetry weighted
into the data for the study. For each of the
asymmetries of interest, the leak-through
systematic is calculated for each of the competing
effects, then added in quadrature to find the total
systematic.

Distortions to the inclusive-jet asymmetry are found to
be negligible. Distortions due to the Collins
effect are reduced by the fact that $\pi^{+}$ and $\pi^{-}$
asymmetries contribute with different signs and similar
magnitudes. For the distortion due to the Collins-like
effect, the asymmetries are modeled to have the same
sign and magnitude for different pion states; however,
the instrumental asymmetries are quite small, greatly
suppressing the amount of possible distortion.

For the distortions to the Collins asymmetry, in particular
due to the Collins-like asymmetry, the shapes are often
quite far from sinusoidal.
However, it is the sinusoidal modulations that determine
the distortions, and these were found to be small in magnitude.
Significant distortions of this kind should also manifest
as highly degraded $\chi^{2}$ distributions for the
asymmetry extractions. As discussed in Section
\ref{subsec:SpinAsymAnalysis}, the distributions are
universally well-behaved, indicating that the present
data are not sensitive to such distortions. Finally, the
gluon-based Collins-like effect should be significant
only at low $p_{T}$, whereas the quark-based Collins
effect should be significant only at high $p_{T}$. Consequently,
no matter the size of the distortion from acceptance, the
final distortions will be highly suppressed by the small
size of the competing asymmetries. Accordingly, the only
possible large distortions would arise from leak-through
of the inclusive-jet asymmetry to the Collins-like asymmetry
at low $p_{T}$, and vice versa. These distortions are highly
suppressed by the relatively uniform acceptance.

A final acceptance-related systematic uncertainty is dilution from
finite azimuthal bin-width. The asymmetry is extracted by fitting
the $\phi$-dependence of the cross-ratios. The inclusive jet
asymmetries are extracted using six azimuthal bins, while the
Collins and Collins-like asymmetries are extracted using twelve
azimuthal bins. The finite size of the bins will introduce a dilution
to the extracted asymmetry, possibly further complicated by acceptance
non-uniformity within the bins. The size of this effect can be calculated
using the infrastructure developed for the leak-through
estimates. Instead of weighting events with input
asymmetries for the competing effects, events are weighted
with input asymmetries for the desired effect. By fitting the resulting
asymmetries, the extracted amplitude gives a precise estimate of
the dilution from finite binning. The extracted value of the dilution
(on the order of $1.5\%$) is consistent with what is expected for
the size of the bins. These dilution values are used as a correction
to the measured asymmetries.

\begin{figure}
\begin{center}
\includegraphics[width=0.49\textwidth]{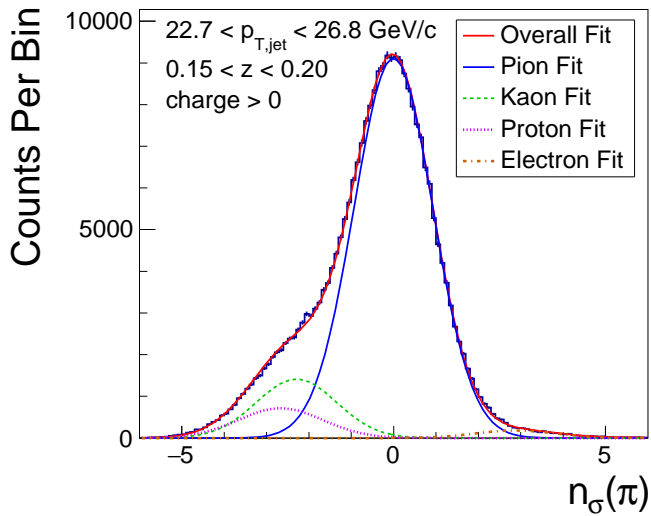}
\end{center}
\caption{\label{fig:Nsig} Multi-gaussian fit to
the distribution of $n_{\sigma}\left(\pi\right)$ for positively
charged particles from jets with $22.7 < p_{T} < 26.8$ GeV$/c$
carrying momentum fractions of $0.15 < z < 0.20$. The overall
fit is shown along with the individual portions for pions, kaons,
protons, and electrons.}
\end{figure}

\subsection{Non-pion Background}
\label{subsec:NonPionBG}

\begin{figure*}
\begin{center}
\includegraphics[width=0.70\textwidth]{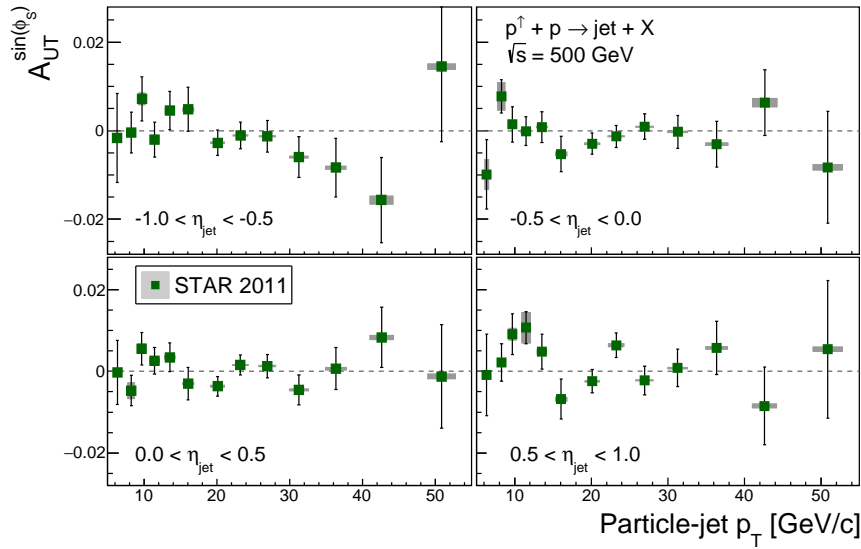}
\end{center}
\caption{\label{fig:SivPt} Inclusive jet asymmetries
$A_{UT}^{\sin(\phi_S)}$, also often referred to as $A_N$, as a function of
particle-jet $p_{T}$. Statistical uncertainties are shown as error bars, while
systematic uncertainties are shown as shaded error boxes. An additional
$3.5\%$ vertical scale uncertainty from polarization is correlated across all
bins. Events are separated into different regions of jet $\eta$, calculated
relative to the polarized beam. Across the full range of jet $p_{T}$, the
measured asymmetries are consistently small.}
\end{figure*}

The backgrounds of main concern for inclusive jets are those
events at the detector-jet level that do not associate with a
parton-jet (Section \ref{subsec:Match}). For pions within jets,
misidentified protons, kaons, and electrons represent an
additional background. To estimate this contamination, data
distributions of $n_{\sigma}(\pi)$ from the TPC (Section
\ref{subsec:PionSelect}) and $m^2$ from TOF \cite{PID} are
analyzed. Signal fractions are extracted by fitting $m^2$ with
a multi-Voigt profile and $n_{\sigma}(\pi)$ with a
multi-Gaussian function to extract yields for pions, kaons,
protons, and electrons over the active ranges. Examples
of these fits are shown in Figs.~\ref{fig:TOF} and \ref{fig:Nsig}.
The $n_{\sigma}(\pi)$ fits utilize corrections to the Bichsel
theoretical $dE/dx$ expectations similar to what is described
in Ref.~\cite{PID_highPt}. For each kinematic bin, the $m^{2}$
fits are performed for three different ranges of $n_{\sigma}(\pi)$,
the values of which are varied according to the kinematics to
create ``pion-rich,'' ``kaon-rich,'' and ``proton-rich'' samples.
The shapes of the Voigt profiles are constrained to be
independent of $n_{\sigma}(\pi)$, while the overall scales
are allowed to vary.

Kaons and pions with a total momentum of 1.1 GeV$/c$
experience the same $dE/dx$ in the STAR TPC (see, e.g.,
Fig.~\ref{fig:TOF}) and protons and pions have the same
$dE/dx$ at a total momentum of 1.7 GeV$/c$. The
multi-Gaussian $n_{\sigma}(\pi)$ fits are insufficient to
determine the non-pion backgrounds in the vicinity of these
crossovers. For bins of jet $p_T$ and pion $z$ corresponding
to pion momenta below $\sim2.1$ GeV$/c$, the TOF $m^{2}$
distributions (e.g., Fig.~\ref{fig:TOF}) are used exclusively to extract
the pion, kaon, and proton fractions. For bins corresponding to
pion momenta from $\sim2.1-3.5$ GeV$/c$, the pion $dE/dx$
value are well separated from those of kaons and protons.
However, kaons and protons experience the same $dE/dx$ at
these kinematics in the STAR TPC. The $m^{2}$ resolution of
the STAR TOF is not sufficient to separate pions from kaons
at these momenta. Thus, a hybrid approach is used. The
$n_{\sigma}(\pi)$ fits are used to extract the pion fractions
and combined kaon and proton fractions. The isolated
proton fractions are extracted from the $m^{2}$ distributions,
where they are still well separated from pions and kaons.
The kaon fractions are then taken from the difference
between the combined $n_{\sigma}(\pi)$ kaon plus proton
fractions and the $m^{2}$ proton fractions. For momenta
above $\sim3.5$ GeV$/c$, pion, kaon, and proton fractions
are extracted exclusively from $n_{\sigma}(\pi)$ fits
(e.g., Fig.~\ref{fig:Nsig}). For all kinematics, electron fractions
are determined from $n_{\sigma}(\pi)$ fits. In the kinematic
regions where the kaon and proton $dE/dx$ values are
similar to those for pions, the kaons and protons contribute
a $\sim10\%$ background. Outside of these kinematic
regions, the kaon and proton backgrounds are tyically in
the range of $1-2\%$. Electrons typically contribute $1\%$
or less.

The extracted pion, kaon, and proton fractions are used to
correct the measured asymmetries. For each kinematic bin,
asymmetries are measured for the three different ranges of
$n_{\sigma}(\pi)$ discussed earlier. The corrected pion
asymmetry is calculated as
\begin{widetext}
\begin{eqnarray}
A_{\pi} & = & \frac{A_{1}\left(f^{K}_{2}f^{p}_{3}-f^{K}_{3}f^{p}_{2}\right)+A_{2}\left(f^{K}_{3}f^{p}_{1}-f^{K}_{1}f^{p}_{3}\right)+A_{3}\left(f^{K}_{1}f^{p}_{2}-f^{K}_{2}f^{p}_{1}\right)}{f^{\pi}_{1}f^{K}_{2}f^{p}_{3}+f^{\pi}_{2}f^{K}_{3}f^{p}_{1}+f^{\pi}_{3}f^{K}_{1}f^{p}_{2}-f^{\pi}_{1}f^{K}_{3}f^{p}_{2}-f^{\pi}_{2}f^{K}_{1}f^{p}_{3}-f^{\pi}_{3}f^{K}_{2}f^{p}_{1}} \mbox{.} \label{eq:BgCorr}
\end{eqnarray}
\end{widetext}
Here, $A_{1}$, $A_{2}$, and $A_{3}$ are the asymmetries
measured in the pion-rich, kaon-rich, and proton-rich $n_{\sigma}(\pi)$
ranges, respectively; $f^{\pi}_{1}$, $f^{K}_{1}$, $f^{p}_{1}$ are,
respectively, the pion, kaon, and proton fractions from the pion-rich
sample; $f^{\pi}_{2}$, $f^{K}_{2}$, and $f^{p}_{2}$ are, respectively, the
pion, kaon, and proton fractions from the kaon-rich sample; and
$f^{\pi}_{3}$, $f^{K}_{3}$, and $f^{p}_{3}$ are, respectively, the
pion, kaon, and proton fractions from the proton-rich sample.
Equation \ref{eq:BgCorr} assumes the electron asymmetry is
negligible. The electron contamination is dominated by photonic
electrons, largely from $\pi^{0}$ decay photons, and heavy flavor
decays. Neutral pion asymmetries are expected to be approximately the averages
of the $\pi^{+}$ and $\pi^{-}$ asymmetries \cite{YuanCollins,*YuanCollinsLong}.
These averages are observed to be small in all cases, indicating the neutral pion
asymmetries should also be small. Heavy flavor is produced primarily through gluon fusion
for the present kinematic range. The $gg \rightarrow q\bar{q}$ process
does not contribute either to the Collins or the Collins-like effect \cite{LeaderSpinBook}.

The systematic uncertainty for the correction is taken
from the full differences between fractions extracted with
different methodologies. For the momentum range below
$\sim2.1$ GeV$/c$, the uncertainties are the differences
between fractions calculated from $n_{\sigma}(\pi)$ fits
and $m^{2}$ fits. For momenta above $\sim2.1$ GeV$/c$, the
uncertainties are the differences between fractions from
$n_{\sigma}(\pi)$ fits with and without fixing function
parameters according to the corrected Bichsel expectations.

\begin{figure*}
\begin{center}
\includegraphics[width=0.465\textwidth]{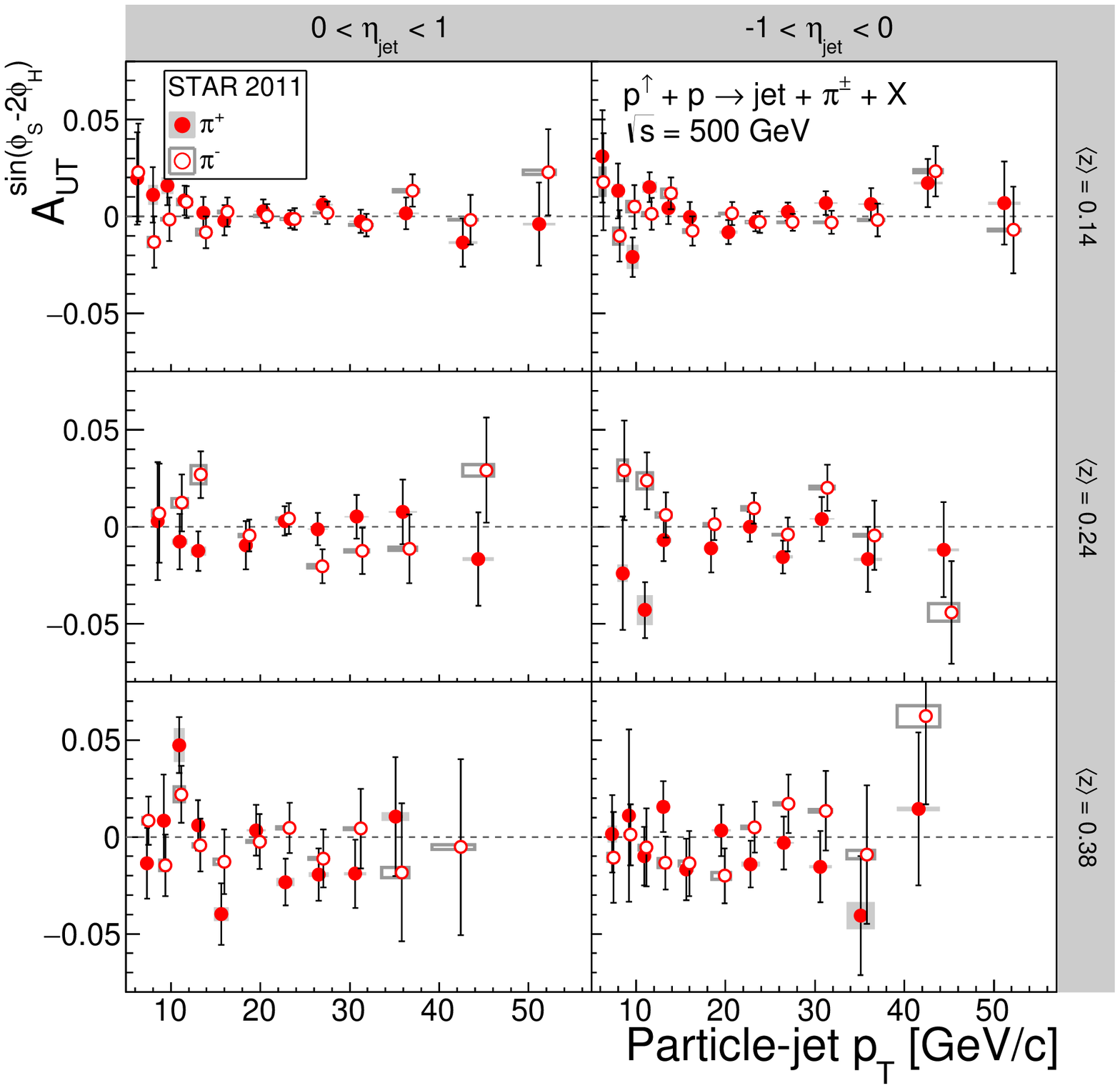}
\includegraphics[width=0.465\textwidth]{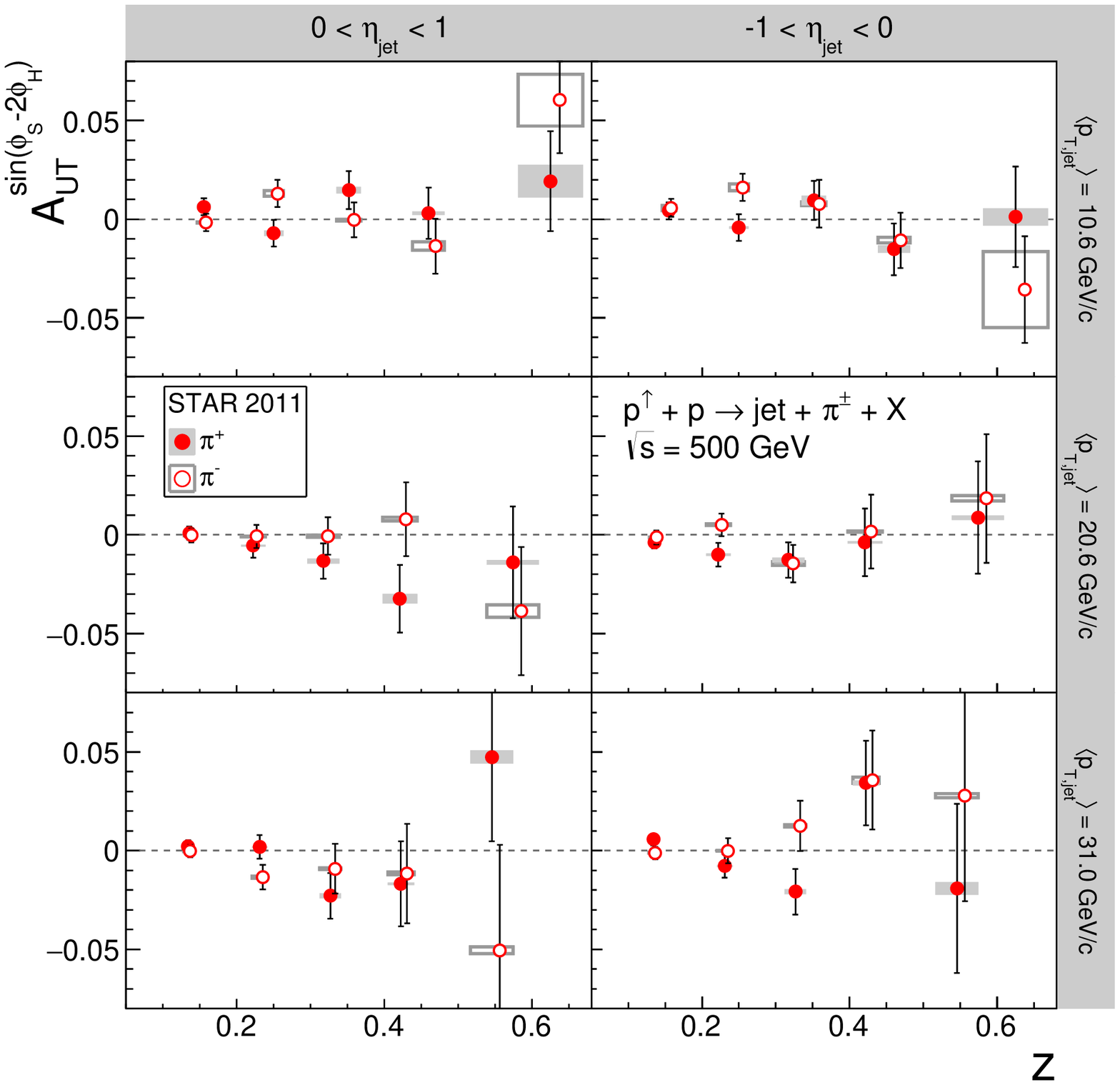}
\end{center}
\caption{\label{fig:ColLike} Collins-like asymmetries as a
function of particle-jet $p_{T}$ (left) and as a function of pion $z$ (right).
$\pi^{-}$ points are shifted horizontally for clarity. Statistical uncertainties
are shown as error bars, while systematic uncertainties are shown as
shaded error boxes.  An additional $3.5\%$ vertical scale uncertainty from
polarization is correlated across all bins. Events are separated into different
regions of jet $\eta$, calculated relative to the polarized beam. The jet $p_{T}$
dependence is shown for three bins of pion $z$: $0.1<z<0.2$, $0.2<z<0.3$,
and $0.3<z<0.8$. The pion $z$ dependence is shown for three bins of
jet $p_{T}$: $6.0 < p_T < 13.8$ GeV/$c$, $13.8 < p_T < 22.7$ GeV/$c$,
and $22.7 < p_T < 55.0$ GeV/$c$. Asymmetries are consistently small.}
\end{figure*}

\subsection{Polarization Uncertainty}

As mentioned in Section \ref{subsec:Experiment}, a correlated
systematic uncertainty of 3.5\% applies to all single-spin
asymmetries, due to estimated uncertainties in the beam
polarization \cite{2011Pol}. This uncertainty includes
a $3.3\%$ contribution from the polarization scale, a $0.9\%$
contribution from fill-to-fill scale uncertainties, and a $0.5\%$
contribution from uncertainties in the transverse polarization
profile correction. Additionally, if the beam polarization is not perfectly
vertical with respect to the laboratory frame, there can also be a systematic uncertainty due to the
horizontal or ``radial'' component of the beam polarization. For
the present analysis, a polarization offset from vertical of $\delta$
will yield a dilution of approximately $1-\delta^{2}/2$, in the limit
of uniform acceptance. The values for $\delta$ are measured by
analyzing the azimuthal dependence of single-spin asymmetries
in the ZDC, which is sensitive to neutral particles such as neutrons
produced close to the beamline. The values for $\delta$ are
generally quite small, typically $\ll0.1$ radians \cite{RadialPol}. Deviations due
to non-uniform acceptance must, then, dominate. Conservatively,
it is estimated that radial polarizations cannot impact the asymmetry
extraction by more than fractions of a percent; and they are therefore ignored.

\section{Results}

\begin{figure*}
\begin{center}
\includegraphics[width=0.49\textwidth]{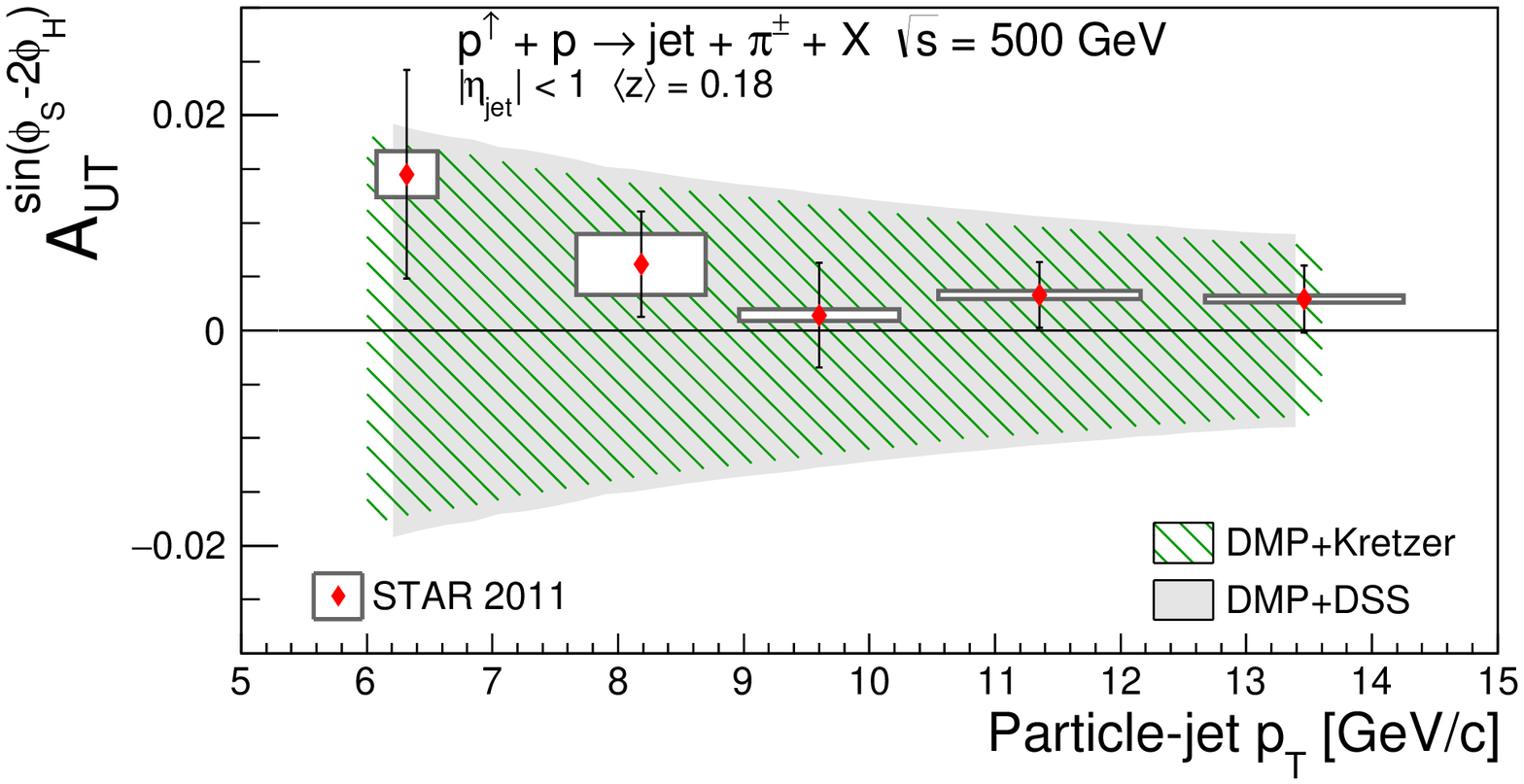}
\includegraphics[width=0.49\textwidth]{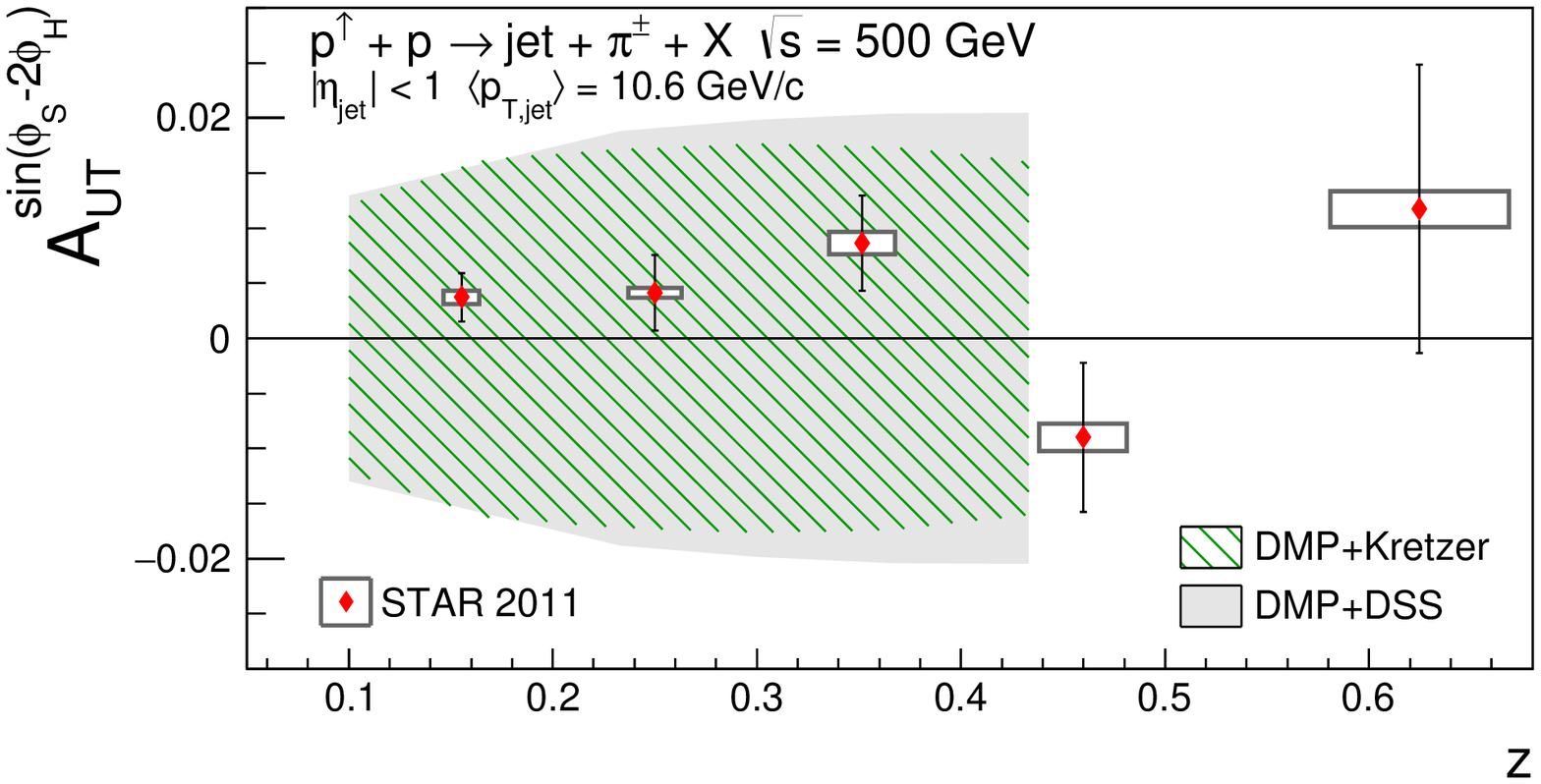}
\end{center}
\caption{\label{fig:ColLikeLow} Collins-like asymmetries as a
function of particle-jet $p_{T}$ for pions reconstructed with $0.1 < z < 0.3$ (left)
and as a function of pion $z$ for jets reconstructed with
$6.0 < p_{T} < 13.8$ GeV$/c$ (right). Asymmetries are shown combining
$\pi^{+}$ and $\pi^{-}$ and integrating over the full range of jet pseudorapidity,
$-1 < \eta < 1$. Statistical uncertainties are shown as error bars, while
systematic uncertainties are shown as shaded error boxes. An additional
$3.5\%$ vertical scale uncertainty from polarization is correlated across all bins.
Shaded bands represent maximal predictions from Ref.~\cite{D'Alesio_2017}
utilizing two sets of fragmentation functions \cite{KretzerFF,DSS}. The
asymmetries are consistently small across the full range of jet $p_{T}$ and
pion $z$ and provide the first experimental constraints on model calculations.}
\end{figure*}

Final results for the inclusive-jet, Collins, and Collins-like
asymmetries are presented in Figs.~\ref{fig:SivPt} through \ref{fig:ColJt}.
In all plots, the statistical uncertainties are shown with error bars,
and systematic uncertainties are shown with shaded error boxes.
The widths of the boxes indicate the total uncertainties in the
kinematic variables that are corrected to the PYTHIA particle-jet
level, as discussed in section \ref{subsec:KineCorr}. For all results, the jet pseudorapidity is calculated relative to
the polarized beam. In the cases of Collins and Collins-like
asymmetries, $\pi^{+}$ is shown with closed circles, $\pi^{-}$ with
open circles, and the combination in closed diamonds. The $3.5\%$
vertical-scale systematic uncertainties from beam polarization are
not shown.

\subsection{Inclusive Jet Asymmetry}
\label{subsec:sivers}

In Fig.~\ref{fig:SivPt} the inclusive jet asymmetries are presented.
The asymmetries are shown as functions of the jet transverse momentum and
presented in separate ranges of jet pseudorapidity, as measured relative to
the polarized beam. A dashed line at zero is provided to guide the eye. In
all cases, the measured asymmetries are consistently small. Integrating over
jet $p_{T}$ and $\eta$ the measured asymmetries are consistent with zero
at the $0.5\sigma$ level with a total uncertainty of $6.0\times10^{-4}$. This is similar
to what has been seen in previous inclusive jet measurements from
$\sqrt{s} = 200$ GeV collisions \cite{STAR_jet_AN2}. The present data
at $\sqrt{s} = 500$ GeV provide stronger limits with better sensitivity to gluonic
subprocesses than previous data at $\sqrt{s} = 200$ GeV \cite{NLO}. Systematic
uncertainties for the present data are well-constrained, with the dominant
uncertainties arising from statistics. The largest systematics arise from
parton-jet matching probabilities at low values of $p_{T}$. At higher values
of $p_{T}$, contributions from leak-through and trigger bias play a more
significant role, though the effects are typically at or below $10\%$ of the
statistical uncertainty.

The inclusive-jet asymmetry is sensitive to the twist-3 distribution
\cite{Efremov_twist3_1,QiuSterman_twist3} related to the $k_{T}$-integrated
Sivers function \cite{AN_Siv,*AN_Siv2,TMDFactBoer}. As the jet
$p_{T}$ increases, the sensitivity to partonic subprocesses changes
(Fig.~\ref{fig:subProcFrac}). At low jet $p_{T}$, the results are more sensitive
to gluonic subprocesses, while sensitivity to quark-based subprocesses
increases at high jet $p_{T}$. Thus the asymmetries at lower values of jet
$p_{T}$ should place constraints on twist-3 PDFs for gluonic interactions
(connected to the gluon Sivers function). As Fig.~\ref{fig:TrigBiasX}
indicates, the lowest jet $p_{T}$ bins are sensitive to
$x_{\mathrm{G}}$ down to $\sim0.01$; while the highest bins probe
$x_{\mathrm{Q}}$ up to $\sim0.2$ for unpolarized $x$.

\subsection{Collins-like Asymmetry}
\label{subsec:colLike}

Results for Collins-like asymmetries are presented in terms of
particle-jet $p_{T}$ and pion $z$ (Figs.~\ref{fig:ColLike} and
\ref{fig:ColLikeLow}). Because the subprocess fraction changes as a
function of particle-jet $p_{T}$, it is informative to examine how the
asymmetries depend on $p_{T}$. The Collins-like asymmetry is
expected to arise from gluon linear polarization \cite{D'Alesio_2011}; thus, the best
sensitivity should reside at lower values of jet $p_{T}$. The left-hand
panel of Fig.~\ref{fig:ColLike} shows the asymmetry as a function of particle-jet
$p_{T}$ for different ranges of jet $\eta$ and pion $z$. The right-hand
panel of Fig.~\ref{fig:ColLike} presents the Collins-like asymmetry
dependence on pion $z$ in bins of jet $\eta$ and jet $p_{T}$.
Across the board, the asymmetries are consistently small. Systematic
uncertainties are well constrained with the dominant uncertainties arising
from statistics. The largest systematics arise from the parton-jet matching
probabilities at low-$p_{T}$ and leak-through at mid-to-high values of $p_{T}$.

From Refs.~\cite{D'Alesio_2017,D'Alesio_2011}, the maximized projections exhibit
the largest asymmetries at lower values of both jet $p_{T}$ and pion $z$.
Furthermore, the maximized projections are similar for $\pi^{+}$ and
$\pi^{-}$ and for $\eta_{\mathrm{jet}}>0$ and $\eta_{\mathrm{jet}}<0$.
Thus, in the left-hand panel of Fig.~\ref{fig:ColLikeLow}, the Collins-like
asymmetries are presented as functions of jet $p_{T}$ for $0.1 < z < 0.3$,
combining pion flavors and integrating over the full range of
$-1 < \eta_{\mathrm{jet}} < 1$. Similarly in the right-hand panel of
Fig.~\ref{fig:ColLikeLow}, the asymmetries are presented as functions
of pion $z$ for jets with $6.0 < p_{T} < 13.8$ GeV$/c$, combining
$\pi^{+}$ and $\pi^{-}$ and integrating over the range
$-1 < \eta_{\mathrm{jet}} < 1$. Again, systematic uncertainties are small,
with the dominant uncertainties arising from statistics. The largest systematics
arise from parton-jet matching probability at low $p_{T}$.

The asymmetries in Fig.~\ref{fig:ColLikeLow} are presented in comparison
with the maximized projections from Ref.~\cite{D'Alesio_2017}. The DMP+Kretzer
calculations utilize fragmentation functions from Ref.~\cite{KretzerFF}, while
the DMP+DSS calculations utilize fragmentation functions from Ref.~\cite{DSS}.
The data place significant constraints on the maximized projections for the
Collins-like asymmetries, representing the first experimental input for this
effect, which is sensitive to linearly polarized gluons.

\begin{figure*}
\begin{center}
\includegraphics[width=0.49\textwidth]{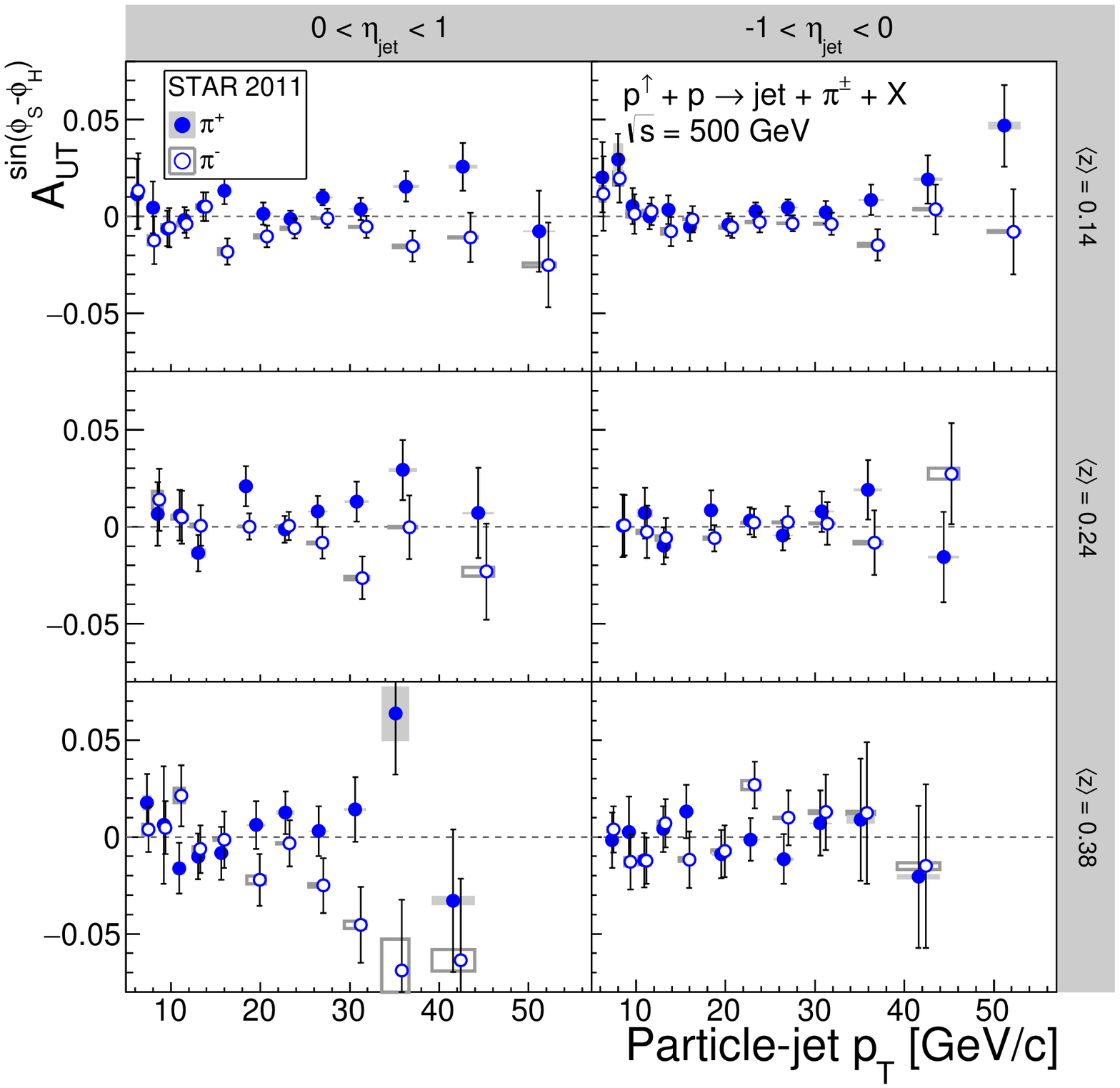}
\includegraphics[width=0.49\textwidth]{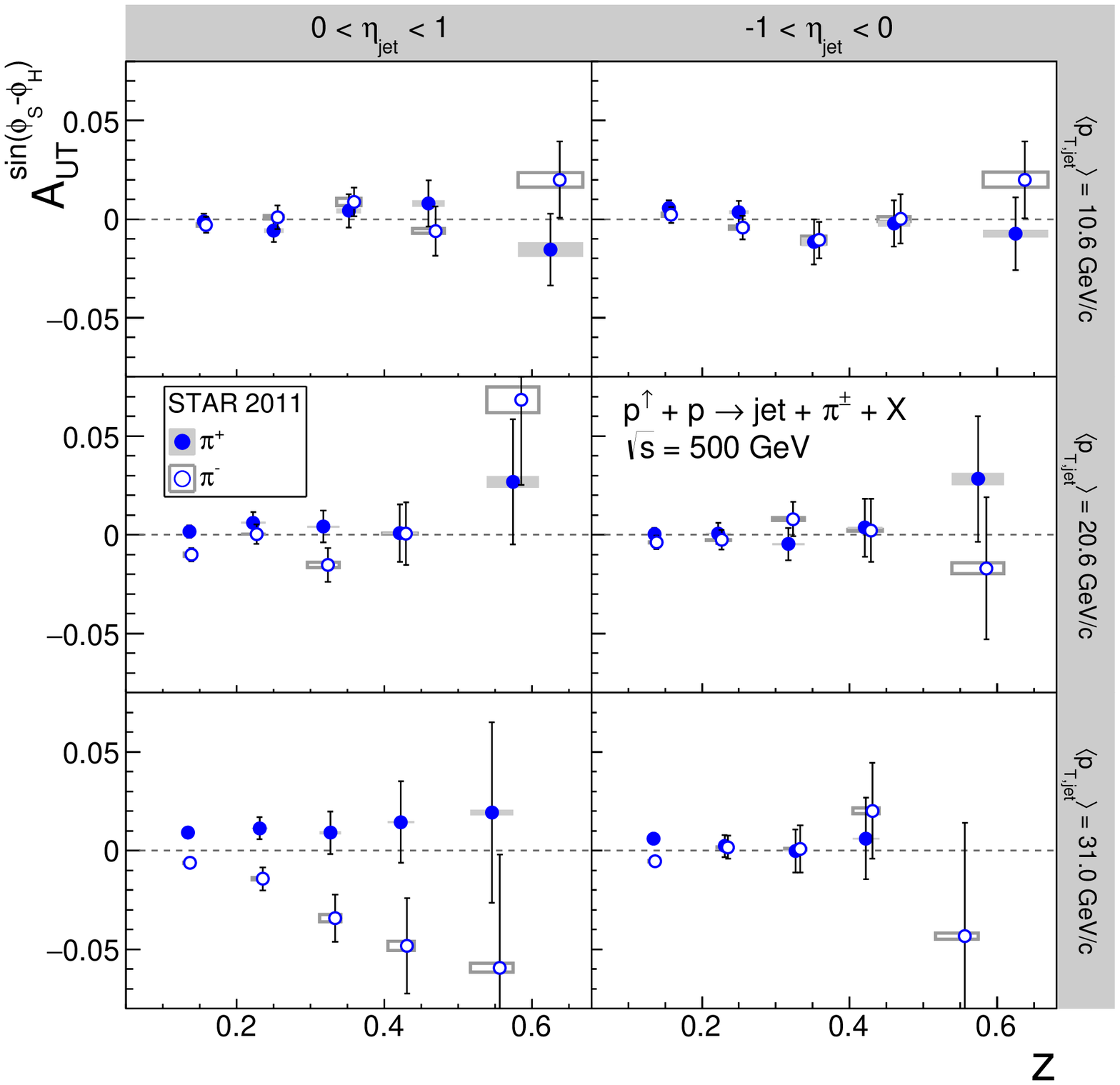}
\end{center}
\caption{\label{fig:Col} Collins asymmetries as a function of
particle-jet $p_{T}$ (left) and as a function of pion $z$ (right). $\pi^{-}$
points are shifted horizontally for clarity. Asymmetries
are shown separately for $\pi^{+}$ and $\pi^{-}$ for two bins of jet $\eta$
(relative to the polarized beam). The jet $p_{T}$ dependence is presented
in three bins of pion $z$: $0.1<z<0.2$, $0.2<z<0.3$, and $0.3<z<0.8$. The
pion $z$ dependence is presented in three bins of jet $p_{T}$: $6.0 < p_T < 13.8$
GeV/$c$, $13.8 < p_T < 22.7$ GeV/$c$, and $22.7 < p_T < 55.0$ GeV/$c$.
Statistical uncertainties are shown as error bars, while systematic uncertainties
are shown as shaded error boxes. An additional $3.5\%$ vertical scale
uncertainty from polarization is correlated across all bins. The asymmetry is
observed to be non-zero for higher values of jet $p_{T}$ and is the first such
observation in polarized proton collisions.}
\end{figure*}

\subsection{Collins Asymmetry}
\label{subsec:col}

\begin{figure}
\begin{center}
\includegraphics[width=0.48\textwidth]{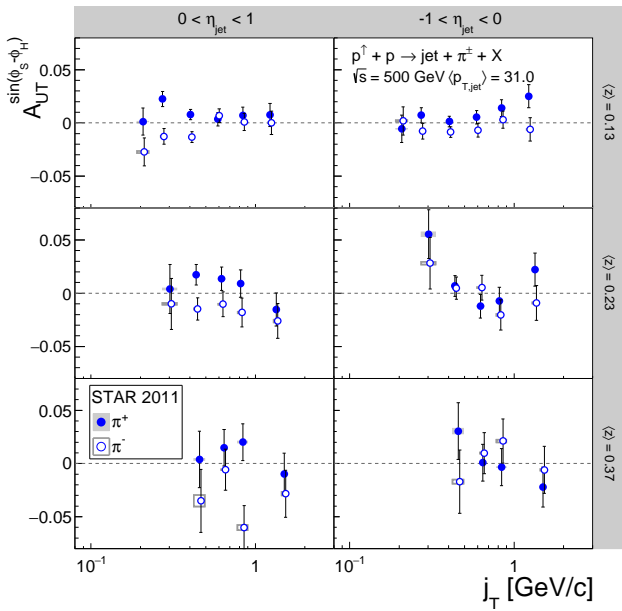}
\end{center}
\caption{\label{fig:ColJt} Collins asymmetries as a function of
pion $j_{T}$ for jets reconstructed with $22.7 < p_{T} < 55.0$ GeV$/c$.
Asymmetries are shown separately for $\pi^{+}$ and $\pi^{-}$ for two bins
of jet $\eta$ (relative to the polarized beam) and three bins of pion $z$:
$0.1<z<0.2$, $0.2<z<0.3$, and $0.3<z<0.8$. Statistical uncertainties are
shown as error bars, while systematic uncertainties are shown as shaded
error boxes. An additional $3.5\%$ vertical scale uncertainty from
polarization is correlated across all bins. Asymmetries appear to show a
dependence upon $j_{T}$ with the largest effects at the lower $j_{T}$ values.}
\end{figure}

Results for Collins asymmetries are presented as functions of
particle-jet $p_{T}$ and pion $z$ in Fig.~\ref{fig:Col} and pion
$j_{T}$ in Fig.~\ref{fig:ColJt}. In contrast to the Collins-like effect,
the Collins asymmetry is expected to arise from quark transversity \cite{AN_Col};
thus, the best sensitivity should reside at higher values of jet $p_{T}$.
In the left-hand panel of Fig.~\ref{fig:Col} the asymmetry is presented
as a function of particle-jet $p_{T}$ for different ranges of jet $\eta$ and
pion $z$. A clear asymmetry is observed in jets with
$p_{T} \gtrsim 20$ GeV$/c$ and $\eta > 0$ relative to the polarized
beam. While statistics are somewhat limited, the magnitude of the
asymmetry also appears to rise from $0.1 < z < 0.2$ to $z > 0.2$.
The observed asymmetries are positive for $\pi^{+}$ and negative
for $\pi^{-}$. Global analyses from SIDIS and $e^{+}e^{-}$ show positive $u$-quark transversity,
negative $d$-quark transversity, positive favored Collins fragmentation functions,
and negative unfavored Collins fragmentation functions. For the present kinematics,
the preponderance of the $\pi^{+}$ and $\pi^{-}$ are expected to materialize from
the favored fragmentation of $u$ and $d$ quarks, respectively. Hence, the
observed charge-dependence is consistent with those of the Collins
asymmetry in SIDIS \cite{HERMESColSiv,*HERMESCol10,COMPASSColSiv05,
*COMPASSColSiv07,*COMPASSColSiv09,*COMPASSColSiv10,*COMPASSCol12,
*COMPASSnew,JLABColSiv} and marks the first such observation in polarized proton collisions.
For pions with $0.1 < z < 0.2$ in jets with $p_{T} \gtrsim 20$ GeV$/c$
and $\eta < 0$ relative to the polarized beam, there are also trends
consistent with a nonzero asymmetry. However, these trends
do not persist for higher values of pion $z$; and thus it is not
clear if this effect is due to physics or simply a statistical fluctuation.

The right-hand panel of Fig.~\ref{fig:Col}, presents the Collins
asymmetries as functions of pion $z$ for three bins of jet $p_{T}$
and two bins of jet $\eta$, calculated relative to the polarized beam.
The magnitude of the Collins asymmetry is expected to change as
a function of $z$ \cite{BELLE2006,*BELLE2012}. At high $p_{T}$
the present data indicate asymmetries with such a dependence.
For jets scattered backward relative to the polarized
beam, as well as for jets with lower values of $p_T$, the measured
asymmetries are small. This is also consistent with
expectation for the Collins effect. As the $p_{T}$ increases, the
present data sample a correspondingly increasing fraction of events from
quark-based partonic subprocesses (Fig.~\ref{fig:subProcFrac})
that are necessary for such effects. Furthermore, the present data
sample higher values of $x$ with increasing values jet $p_{T}$. Current
global extractions of transversity indicate a potentially strong
$x$-dependence \cite{Transversity07,Transversity09,Transversity13,Transversity15,TransversityKang}.
Unpolarized Monte Carlo simulations indicate that for jets reconstructed
with $22.7 < p_{T} < 55.0$ GeV$/c$, the sampled quark $x$ range peaks
around $x\sim0.15$ (Fig.~\ref{fig:TrigBiasX}). This corresponds to the
region where the largest values of transversity are expected
\cite{Transversity07,Transversity09,Transversity13,Transversity15,TransversityKang}
but with $Q^{2}\approx960$ GeV$^{2}$, significantly higher than the
scale probed by the SIDIS data with $Q^{2} < 20$ GeV$^{2}$.

For the present data, systematic uncertainties are small compared
to the statistical uncertainties. As with the Collins-like asymmetries,
the dominant systematics at low jet $p_{T}$
arise from the parton-jet matching probabilities. At higher jet $p_{T}$
values, the dominant sources of systematics arise from leak-through
and trigger bias, though the systematic uncertainties are typically only
$\sim15\%$ of the statistical uncertainties.

The Collins effect requires not only non-zero transversity but also
the presence of a polarized and transverse-momentum-dependent
fragmentation function. Thus, while it is informative to examine the
jet $p_{T}$ and pion $z$ dependences of the Collins effect, it is also
important to examine its $j_{T}$ dependence.
This study examines how the asymmetry depends upon the relative
transverse momentum of the pion, in other words, the pion momentum
transverse to the jet axis. The results for the present data are presented
in Fig.~\ref{fig:ColJt} for jets with $22.7 < p_{T} < 55.0$ GeV$/c$ in three
bins of pion $z$. The asymmetries appear largest around
$j_{T} \sim 0.3-0.4$ GeV$/c$. It is worth noting, again, that the choice
of a lower limit on $\Delta R$ restricts the lower reach of $j_{T}$, in
particular at higher values of pion $z$ or jet $p_{T}$ (Eq.~\ref{eqn:RminJt}).

Integrating over all bins of $z$ at high jet $p_{T}$, the present Collins
asymmetries for $\pi^{+}$ and $\pi^{-}$ are found to be different with a
significance of greater than $5.3\sigma$. Consequently, the present
data represent the first observation of the Collins effect in polarized-proton
collisions.

\subsection{Comparison to Models}
The present data span a range of quark $x$ which complements existing SIDIS
measurements and current transversity extractions
\cite{Transversity07,Transversity09,Transversity13,Transversity15,TransversityKang}
but at one-to-two orders of magnitude higher in $Q^{2}$, as seen in Fig.~\ref{fig:x-q2}.
Accordingly, the present data present an opportunity to address existing
theoretical questions concerning universality and the size of possible
TMD factorization-breaking effects in polarized-proton collisions,
and TMD evolution.

\begin{figure*}
\begin{center}
\includegraphics[width=0.7\textwidth]{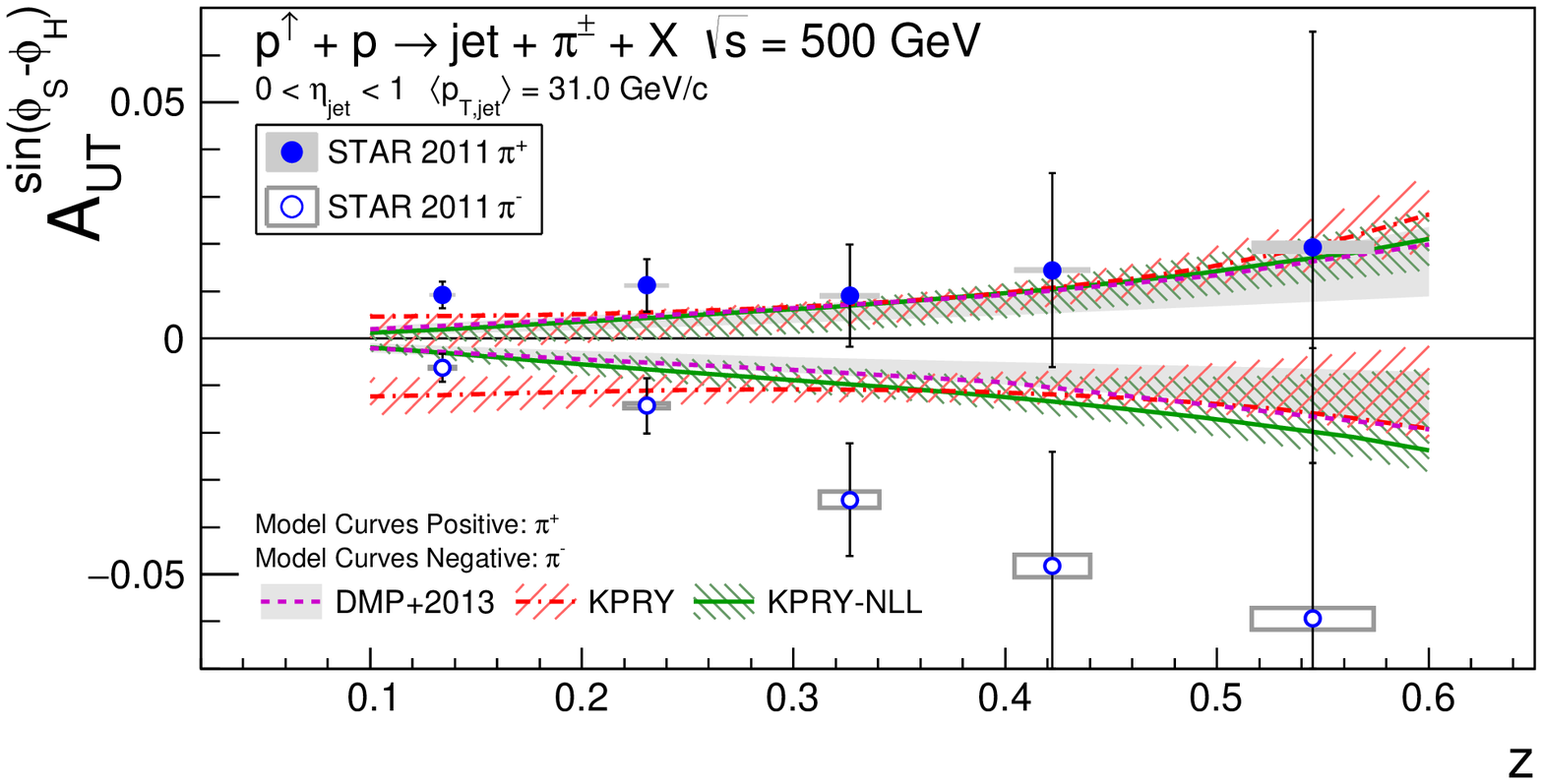}
\end{center}
\caption{\label{fig:ColModels} Collins asymmetries as a
function of pion $z$ for jets reconstructed with $22.7 < p_{T} < 55.0$
GeV$/c$ and $0 < \eta < 1$. The asymmetries are shown in
comparison with model calculations from
Refs.~\cite{D'Alesio_2017,KPRY}. The calculations are based upon
SIDIS and $e^{+}e^{-}$ results and assume robust factorization and
universality of the Collins function. The DMP+2013 \cite{D'Alesio_2017}
and KPRY \cite{KPRY} predictions assume no TMD evolution, while
the KPRY-NLL \cite{KPRY} curves assume TMD evolution up to
next-to-leading-log. All predictions are shown with shaded bands
corresponding to the size of their associated theoretical uncertainties.
The general agreement between the data and the model calculations
is consistent with assumptions of robust TMD-factorization and
universality of the Collins function.}
\end{figure*}

Figure \ref{fig:ColModels} presents the Collins asymmetries for jets
reconstructed with $22.7 < p_{T} < 55.0$ GeV$/c$ and $0 < \eta < 1$
in comparison with three sets of model calculations. Each set is
based upon a global analysis of SIDIS and $e^{+}e^{-}$ data,
assumes robust TMD factorization applied to proton-proton
interactions, and assumes universality of the Collins fragmentation
function. The DMP+2013 predictions are based upon
Refs.~\cite{D'Alesio_2017,Transversity13,D'Alesio_2011} and
utilize the fragmentation functions from Ref.~\cite{DSS}.
Reference \cite{D'Alesio_2017} also presents predictions based
upon older models from Refs.~\cite{Transversity07,Transversity09}.
These older predictions demonstrate that the size of the expected
asymmetries can be sensitive to the choice of fragmentation functions,
in particular for the $g\to\pi^{\pm}$ contribution that still has considerable
uncertainty \cite{DSS14}. The KPRY and KPRY-NLL
predictions are based upon Ref.~\cite{KPRY}. The KPRY-NLL curves
assume TMD evolution up to next-to-leading-log, while the KPRY
curves assume no TMD evolution. In general the models compare
favorably with the data, consistent with the expectation of universality
of the Collins fragmentation function. In addition, this comparison
is also consistent with the assumption of robust TMD factorization
for proton-proton interactions. While it is generally expected that
TMD factorization is broken for proton-proton interactions, it has
been argued that such factorization holds for observation of a
hadron fragment within a jet \cite{TMDFF,YuanCollins,*YuanCollinsLong}. Within theoretical
uncertainties, the data agree relatively well with either assumption
of TMD evolution from the KPRY predictions. However, the data
do show a slight preference for the model without TMD evolution
($\chi^{2} = 14.0$ for $10$ degrees of freedom without evolution
compared with $\chi^{2} = 17.6$ with evolution, using the data
statistical and systematic uncertainties). The measured asymmetries
are generally larger in magnitude than the model predictions, in
particular for $\pi^{-}$. A $\chi^{2}$ test indicates the measurement
and predictions are consistent at the $95\%$ confidence level.

Finally, it is worth noting that polarized-proton collisions
at STAR have also yielded non-zero asymmetries sensitive to transversity
through dihadron interference fragmentation functions
\cite{STAR_IFF2006,*STAR_IFF2011}. These asymmetries persist in
the collinear framework of pQCD, where factorization and universality
are expected to hold \cite{BacchettaRadici}. Efforts to include these
results in global analyses aimed at extracting transversity have already begun \cite{RadiciDIS2017}.
The combination of the present results with those from $e^{+}e^{-}$, SIDIS,
and dihadrons from $p+p$, provide the opportunity for a comprehensive global
analysis to address questions concerning TMD-factorization-breaking, universality,
and evolution.

\section{Conclusions}
We have reported the first measurements of transverse single-spin
asymmetries from inclusive jet and $\mathrm{jet}+\pi^{\pm}$ production
in the central pseudorapidity range from $p^{\uparrow}+p$ at
$\sqrt{s} = 500$ GeV. The data were collected in  2011 with the STAR
detector. As in previous measurements at 200 GeV, the inclusive jet
asymmetry is consistent with zero at the available precision. The first-ever
measurements of the ``Collins-like'' asymmetry, sensitive to linearly
polarized gluons in a polarized proton, are found to be small and provide the first
constraints on model calculations. For the first time, we observe a
non-zero Collins asymmetry in polarized-proton collisions. The data
probe values of $Q^{2}$ significantly higher than existing measurements
from SIDIS. The asymmetries
exhibit a dependence on pion $z$ and are consistent in magnitude for
the two charged-pion species. For $\pi^{+}$, asymmetries are found to
be positive; while those for $\pi^{-}$ are found to be negative. The
present data are compared to Collins asymmetry predictions based
upon SIDIS and $e^{+}e^{-}$ data. The comparisons are consistent
with the expectation for TMD factorization in proton-proton
collisions and universality of the Collins fragmentation function. The
data show a slight preference for models assuming no suppression
from TMD evolution. Further insight into these theoretical questions
can be gained from a global analysis, including dihadron asymmetries
and Collins asymmetries from STAR.

\begin{acknowledgments}
%version dated 2016-10-24 13:50
%retrieved August 22, 2017.
The authors thank Zhong-Bo Kang, Alexei Prokudin, Umberto
D'Alesio, and Cristian Pisano for valuable discussions and providing the results of their
model calculations.
We thank the RHIC Operations Group and RCF at BNL, the NERSC Center at LBNL,
and the Open Science Grid consortium for providing resources and support. This work
was supported in part by the Office of Nuclear Physics within the U.S.~DOE Office of
Science, the U.S.~National Science Foundation, the Ministry of Education and Science
of the Russian Federation, National Natural Science Foundation of China, Chinese
Academy of Science, the Ministry of Science and Technology of China and the
Chinese Ministry of Education, the National Research Foundation of Korea, GA and
MSMT of the Czech Republic, Department of Atomic Energy and Department of
Science and Technology of the Government of India; the National Science Centre of
Poland, National Research Foundation, the Ministry of Science, Education and Sports
of the Republic of Croatia, RosAtom of Russia and German Bundesministerium fur
Bildung, Wissenschaft, Forschung and Technologie (BMBF) and the Helmholtz Association.

\end{acknowledgments}

\bibliography{2011JetPaper.v5.1}% Produces the bibliography via BibTeX.

\begin{thebibliography}{108}
\expandafter\ifx\csname natexlab\endcsname\relax\def\natexlab#1{#1}\fi
\expandafter\ifx\csname bibnamefont\endcsname\relax
  \def\bibnamefont#1{#1}\fi
\expandafter\ifx\csname bibfnamefont\endcsname\relax
  \def\bibfnamefont#1{#1}\fi
\expandafter\ifx\csname citenamefont\endcsname\relax
  \def\citenamefont#1{#1}\fi
\expandafter\ifx\csname url\endcsname\relax
  \def\url#1{\texttt{#1}}\fi
\expandafter\ifx\csname urlprefix\endcsname\relax\def\urlprefix{URL }\fi
\providecommand{\bibinfo}[2]{#2}
\providecommand{\eprint}[2][]{\url{#2}}

\bibitem[{\citenamefont{Ralston and Soper}(1979)}]{RalstonSoper}
\bibinfo{author}{\bibfnamefont{J.}~\bibnamefont{Ralston}} \bibnamefont{and}
  \bibinfo{author}{\bibfnamefont{D.}~\bibnamefont{Soper}},
  \bibinfo{journal}{Nucl. Phys. B} \textbf{\bibinfo{volume}{152}},
  \bibinfo{pages}{109} (\bibinfo{year}{1979});
\bibitem[{\citenamefont{Jaffe and Ji}(1991)}]{JaffeJi}
\bibinfo{author}{\bibfnamefont{R.~L.} \bibnamefont{Jaffe}} \bibnamefont{and}
  \bibinfo{author}{\bibfnamefont{X.}~\bibnamefont{Ji}}, \bibinfo{journal}{Phys.
  Rev. Lett.} \textbf{\bibinfo{volume}{67}}, \bibinfo{pages}{552}
  (\bibinfo{year}{1991}).

\bibitem[{\citenamefont{Aaron et~al.}(2010)}]{HERA}
\bibinfo{author}{\bibfnamefont{F.~D.} \bibnamefont{Aaron}} \bibnamefont{et~al.}
  (\bibinfo{collaboration}{H1 and ZEUS Collaborations}),
  \bibinfo{journal}{JHEP} \textbf{\bibinfo{volume}{1001}}, \bibinfo{pages}{109}
  (\bibinfo{year}{2010}).

\bibitem[{\citenamefont{de~Florian et~al.}(2008)\citenamefont{de~Florian,
  Sassot, Stratmann, and Vogelsang}}]{DSSV08}
\bibinfo{author}{\bibfnamefont{D.}~\bibnamefont{de~Florian}},
  \bibinfo{author}{\bibfnamefont{R.}~\bibnamefont{Sassot}},
  \bibinfo{author}{\bibfnamefont{M.}~\bibnamefont{Stratmann}},
  \bibnamefont{and}
  \bibinfo{author}{\bibfnamefont{W.}~\bibnamefont{Vogelsang}},
  \bibinfo{journal}{Phys. Rev. Lett.} \textbf{\bibinfo{volume}{101}},
  \bibinfo{pages}{072001} (\bibinfo{year}{2008});
\bibitem[{\citenamefont{de~Florian et~al.}(2009)\citenamefont{de~Florian,
  Sassot, Stratmann, and Vogelsang}}]{DSSV09}
  \bibinfo{journal}{Phys. Rev. D} \textbf{\bibinfo{volume}{80}},
  \bibinfo{pages}{034030} (\bibinfo{year}{2009});
\bibitem[{\citenamefont{de~Florian et~al.}(2014)\citenamefont{de~Florian,
  Sassot, Stratmann, and Vogelsang}}]{DSSV14}
  \bibinfo{journal}{Phys. Rev. Lett.} \textbf{\bibinfo{volume}{113}},
  \bibinfo{pages}{012001} (\bibinfo{year}{2014}).


\bibitem[{\citenamefont{Bl\"{u}mlein and B\"{o}ttcher}(2010)}]{BB}
\bibinfo{author}{\bibfnamefont{J.}~\bibnamefont{Bl\"{u}mlein}}
  \bibnamefont{and}
  \bibinfo{author}{\bibfnamefont{H.}~\bibnamefont{B\"{o}ttcher}},
  \bibinfo{journal}{Nucl. Phys.} \textbf{\bibinfo{volume}{B841}},
  \bibinfo{pages}{205} (\bibinfo{year}{2010}).

\bibitem[{\citenamefont{Leader et~al.}(2010)\citenamefont{Leader, Sidorov, and
  Stamenov}}]{LSS}
\bibinfo{author}{\bibfnamefont{E.}~\bibnamefont{Leader}},
  \bibinfo{author}{\bibfnamefont{A.~V.} \bibnamefont{Sidorov}},
  \bibnamefont{and} \bibinfo{author}{\bibfnamefont{D.~B.}
  \bibnamefont{Stamenov}}, \bibinfo{journal}{Phys. Rev. D}
  \textbf{\bibinfo{volume}{82}}, \bibinfo{pages}{114018}
  (\bibinfo{year}{2010}).

\bibitem[{\citenamefont{Ball et~al.}(2013)}]{NNPDF13}
\bibinfo{author}{\bibfnamefont{R.~D.} \bibnamefont{Ball}} \bibnamefont{et~al.}
  (\bibinfo{collaboration}{NNPDF Collaboration}), \bibinfo{journal}{Nucl.
  Phys.} \textbf{\bibinfo{volume}{B874}}, \bibinfo{pages}{36}
  (\bibinfo{year}{2013}); \bibitem[{\citenamefont{Nocera et~al.}(2014)}]{NNPDF14}
\bibinfo{author}{\bibfnamefont{E.~R.} \bibnamefont{Nocera}}
  \bibnamefont{et~al.} (\bibinfo{collaboration}{NNPDF Collaboration}),
  \bibinfo{journal}{\textit{ibid.}} \textbf{\bibinfo{volume}{B887}},
  \bibinfo{pages}{276} (\bibinfo{year}{2014}).

\bibitem[{\citenamefont{Kane et~al.}(1978)\citenamefont{Kane, Pumplin, and
  Repko}}]{KPR}
\bibinfo{author}{\bibfnamefont{G.~L.}~\bibnamefont{Kane}},
  \bibinfo{author}{\bibfnamefont{J.}~\bibnamefont{Pumplin}}, \bibnamefont{and}
  \bibinfo{author}{\bibfnamefont{W.}~\bibnamefont{Repko}},
  \bibinfo{journal}{Phys.~Rev.~Lett.} \textbf{\bibinfo{volume}{41}},
  \bibinfo{pages}{1689} (\bibinfo{year}{1978}).

\bibitem[{\citenamefont{Dragoset et~al.}(1978)\citenamefont{Dragoset, Roberts,
  Bowers, Courant, Kagan et~al.}}]{ZGSAN}
\bibinfo{author}{\bibfnamefont{W.}~\bibnamefont{Dragoset}},
  \bibinfo{author}{\bibfnamefont{J.}~\bibnamefont{Roberts}},
  \bibinfo{author}{\bibfnamefont{J.}~\bibnamefont{Bowers}},
  \bibinfo{author}{\bibfnamefont{H.}~\bibnamefont{Courant}},
  \bibinfo{author}{\bibfnamefont{H.}~\bibnamefont{Kagan}},
  \bibnamefont{et~al.}, \bibinfo{journal}{Phys. Rev. D}
  \textbf{\bibinfo{volume}{18}}, \bibinfo{pages}{3939} (\bibinfo{year}{1978}).

\bibitem[{\citenamefont{Antille et~al.}(1980)\citenamefont{Antille, Dick,
  Madansky, Perret-Gallix, Werlen et~al.}}]{CERNPSAN}
\bibinfo{author}{\bibfnamefont{J.}~\bibnamefont{Antille}},
  \bibinfo{author}{\bibfnamefont{L.}~\bibnamefont{Dick}},
  \bibinfo{author}{\bibfnamefont{L.}~\bibnamefont{Madansky}},
  \bibinfo{author}{\bibfnamefont{D.}~\bibnamefont{Perret-Gallix}},
  \bibinfo{author}{\bibfnamefont{M.}~\bibnamefont{Werlen}},
  \bibnamefont{et~al.}, \bibinfo{journal}{Phys. Lett. B}
  \textbf{\bibinfo{volume}{94}}, \bibinfo{pages}{523} (\bibinfo{year}{1980}).

\bibitem[{\citenamefont{Bonner et~al.}(1988)}]{E704-88}
\bibinfo{author}{\bibfnamefont{B.~E.} \bibnamefont{Bonner}}
  \bibnamefont{et~al.} (\bibinfo{collaboration}{FNAL-E704 Collaboration}),
  \bibinfo{journal}{Phys. Rev. Lett.} \textbf{\bibinfo{volume}{61}},
  \bibinfo{pages}{1918} (\bibinfo{year}{1988});
  \bibitem[{\citenamefont{Adams et~al.}(1991{\natexlab{a}})}]{E704}
\bibinfo{author}{\bibfnamefont{D.~L.} \bibnamefont{Adams}} \bibnamefont{et~al.}
  (\bibinfo{collaboration}{FNAL E581/704 Collaboration}),
  \bibinfo{journal}{Phys. Lett. B} \textbf{\bibinfo{volume}{261}},
  \bibinfo{pages}{201} (\bibinfo{year}{1991}{\natexlab{a}});
\bibitem[{\citenamefont{Adams et~al.}(1991{\natexlab{b}})}]{E704chg}
\bibinfo{author}{\bibfnamefont{D.~L.} \bibnamefont{Adams}} \bibnamefont{et~al.}
  (\bibinfo{collaboration}{FNAL E704 Collaboration}), \bibinfo{journal}{\textit{ibid.}} \textbf{\bibinfo{volume}{264}}, \bibinfo{pages}{462}
  (\bibinfo{year}{1991}{\natexlab{b}});
  \bibitem[{\citenamefont{Bravar et~al.}(1996)}]{E704chgpbar}
\bibinfo{author}{\bibfnamefont{A.}~\bibnamefont{Bravar}} \bibnamefont{et~al.}
  (\bibinfo{collaboration}{Fermilab E704 Collaboration}),
  \bibinfo{journal}{Phys. Rev. Lett.} \textbf{\bibinfo{volume}{77}},
  \bibinfo{pages}{2626} (\bibinfo{year}{1996}).

\bibitem[{\citenamefont{Saroff et~al.}(1990)\citenamefont{Saroff, Baller,
  Blazey, Courant, Heller et~al.}}]{AGSANchg}
\bibinfo{author}{\bibfnamefont{S.}~\bibnamefont{Saroff}},
  \bibinfo{author}{\bibfnamefont{B.}~\bibnamefont{Baller}},
  \bibinfo{author}{\bibfnamefont{G.}~\bibnamefont{Blazey}},
  \bibinfo{author}{\bibfnamefont{H.}~\bibnamefont{Courant}},
  \bibinfo{author}{\bibfnamefont{K.~J.} \bibnamefont{Heller}},
  \bibnamefont{et~al.}, \bibinfo{journal}{Phys. Rev. Lett.}
  \textbf{\bibinfo{volume}{64}}, \bibinfo{pages}{995} (\bibinfo{year}{1990}).

\bibitem[{\citenamefont{Adams et~al.}(2004)}]{STAR_FPD_AN1}
\bibinfo{author}{\bibfnamefont{J.}~\bibnamefont{Adams}} \bibnamefont{et~al.}
  (\bibinfo{collaboration}{STAR Collaboration}), \bibinfo{journal}{Phys. Rev.
  Lett.} \textbf{\bibinfo{volume}{92}}, \bibinfo{pages}{171801}
  (\bibinfo{year}{2004});
\bibitem[{\citenamefont{Abelev et~al.}(2008)}]{STAR_FPD_AN2}
\bibinfo{author}{\bibfnamefont{B.~I.} \bibnamefont{Abelev}}
  \bibnamefont{et~al.} (\bibinfo{collaboration}{STAR Collaboration}),
  \bibinfo{journal}{\textit{ibid.}} \textbf{\bibinfo{volume}{101}},
  \bibinfo{pages}{222001} (\bibinfo{year}{2008});
\bibitem[{\citenamefont{Adamczyk et~al.}(2012{\natexlab{a}})}]{STAR_FPD2_xSec}
\bibinfo{author}{\bibfnamefont{L.}~\bibnamefont{Adamczyk}} \bibnamefont{et~al.}
  (\bibinfo{collaboration}{STAR Collaboration}), \bibinfo{journal}{Phys. Rev.
  D} \textbf{\bibinfo{volume}{86}}, \bibinfo{pages}{051101(R)}
  (\bibinfo{year}{2012}{\natexlab{a}}).

\bibitem[{\citenamefont{Arsene et~al.}(2008)\citenamefont{Arsene, Bearden,
  Beavis, Bekele, Besliu, Budick, B\o{}ggild, Chasman, Dalsgaard, Debbe
  et~al.}}]{BRAHMSAN}
\bibinfo{author}{\bibfnamefont{I.}~\bibnamefont{Arsene}},
  \bibinfo{author}{\bibfnamefont{I.~G.} \bibnamefont{Bearden}},
  \bibinfo{author}{\bibfnamefont{D.}~\bibnamefont{Beavis}},
  \bibinfo{author}{\bibfnamefont{S.}~\bibnamefont{Bekele}},
  \bibinfo{author}{\bibfnamefont{C.}~\bibnamefont{Besliu}},
  \bibinfo{author}{\bibfnamefont{B.}~\bibnamefont{Budick}},
  \bibinfo{author}{\bibfnamefont{H.}~\bibnamefont{B\o{}ggild}},
  \bibinfo{author}{\bibfnamefont{C.}~\bibnamefont{Chasman}},
  \bibinfo{author}{\bibfnamefont{H.~H.} \bibnamefont{Dalsgaard}},
  \bibinfo{author}{\bibfnamefont{R.}~\bibnamefont{Debbe}}, \bibnamefont{et~al.}
  (\bibinfo{collaboration}{BRAHMS Collaboration}), \bibinfo{journal}{Phys. Rev.
  Lett.} \textbf{\bibinfo{volume}{101}}, \bibinfo{pages}{042001}
  (\bibinfo{year}{2008}).

\bibitem[{\citenamefont{Adare et~al.}(2014)}]{PHENIXMPCAN}
\bibinfo{author}{\bibfnamefont{A.}~\bibnamefont{Adare}} \bibnamefont{et~al.}
  (\bibinfo{collaboration}{PHENIX Collaboration}), \bibinfo{journal}{Phys. Rev.
  D} \textbf{\bibinfo{volume}{90}}, \bibinfo{pages}{012006}
  (\bibinfo{year}{2014}).

\bibitem[{\citenamefont{Efremov and Teryaev}(1982)}]{Efremov_twist3_1}
\bibinfo{author}{\bibfnamefont{A.}~\bibnamefont{Efremov}} \bibnamefont{and}
  \bibinfo{author}{\bibfnamefont{O.}~\bibnamefont{Teryaev}},
  \bibinfo{journal}{Yad.~Fiz.} \textbf{\bibinfo{volume}{36}},
  \bibinfo{pages}{242} (\bibinfo{year}{1982}),
  \bibinfo{note}{[Sov.~J.~Nucl.~Phys.~36, 140 (1982)]}.

\bibitem[{\citenamefont{Qiu and Sterman}(1998)}]{QiuSterman_twist3}
\bibinfo{author}{\bibfnamefont{J.}~\bibnamefont{Qiu}} \bibnamefont{and}
  \bibinfo{author}{\bibfnamefont{G.}~\bibnamefont{Sterman}},
  \bibinfo{journal}{Phys. Rev. D} \textbf{\bibinfo{volume}{59}},
  \bibinfo{pages}{014004} (\bibinfo{year}{1998}).

\bibitem[{\citenamefont{Kanazawa and Koike}(2013)}]{KanazawaJet}
\bibinfo{author}{\bibfnamefont{K.}~\bibnamefont{Kanazawa}} \bibnamefont{and}
  \bibinfo{author}{\bibfnamefont{Y.}~\bibnamefont{Koike}},
  \bibinfo{journal}{Phys. Lett. B} \textbf{\bibinfo{volume}{720}},
  \bibinfo{pages}{161} (\bibinfo{year}{2013}).

\bibitem[{\citenamefont{Kanazawa et~al.}(2014)\citenamefont{Kanazawa, Koike,
  Metz, and Pitonyak}}]{Pitonyak}
\bibinfo{author}{\bibfnamefont{K.}~\bibnamefont{Kanazawa}},
  \bibinfo{author}{\bibfnamefont{Y.}~\bibnamefont{Koike}},
  \bibinfo{author}{\bibfnamefont{A.}~\bibnamefont{Metz}}, \bibnamefont{and}
  \bibinfo{author}{\bibfnamefont{D.}~\bibnamefont{Pitonyak}},
  \bibinfo{journal}{Phys. Rev. D} \textbf{\bibinfo{volume}{89}},
  \bibinfo{pages}{111501(R)} (\bibinfo{year}{2014}).

\bibitem[{\citenamefont{Sivers}(1990)}]{AN_Siv}
\bibinfo{author}{\bibfnamefont{D.}~\bibnamefont{Sivers}},
  \bibinfo{journal}{Phys. Rev. D} \textbf{\bibinfo{volume}{41}},
  \bibinfo{pages}{83} (\bibinfo{year}{1990});
\bibitem[{\citenamefont{Sivers}(1991)}]{AN_Siv2}
 \textbf{\bibinfo{volume}{43}},
  \bibinfo{pages}{261} (\bibinfo{year}{1991}).

\bibitem[{\citenamefont{Collins}(1993)}]{AN_Col}
\bibinfo{author}{\bibfnamefont{J.}~\bibnamefont{Collins}},
  \bibinfo{journal}{Nucl. Phys.} \textbf{\bibinfo{volume}{B396}},
  \bibinfo{pages}{161} (\bibinfo{year}{1993}).

\bibitem[{\citenamefont{Brodsky et~al.}(2002)\citenamefont{Brodsky, Hwang, and
  Schmidt}}]{BrodHwangSchmidt}
\bibinfo{author}{\bibfnamefont{S.~J.} \bibnamefont{Brodsky}},
  \bibinfo{author}{\bibfnamefont{D.~S.} \bibnamefont{Hwang}}, \bibnamefont{and}
  \bibinfo{author}{\bibfnamefont{I.}~\bibnamefont{Schmidt}},
  \bibinfo{journal}{Phys. Lett. B} \textbf{\bibinfo{volume}{530}},
  \bibinfo{pages}{99} (\bibinfo{year}{2002}).

\bibitem[{\citenamefont{Collins}(2002)}]{DYFactCollins}
\bibinfo{author}{\bibfnamefont{J.~C.} \bibnamefont{Collins}},
  \bibinfo{journal}{Phys. Lett. B} \textbf{\bibinfo{volume}{536}},
  \bibinfo{pages}{43} (\bibinfo{year}{2002}).

\bibitem[{\citenamefont{Boer et~al.}(2003)\citenamefont{Boer, Mulders, and
  Pijlman}}]{TMDFactBoer}
\bibinfo{author}{\bibfnamefont{D.}~\bibnamefont{Boer}},
  \bibinfo{author}{\bibfnamefont{P.~J.} \bibnamefont{Mulders}},
  \bibnamefont{and} \bibinfo{author}{\bibfnamefont{F.}~\bibnamefont{Pijlman}},
  \bibinfo{journal}{Nucl. Phys.} \textbf{\bibinfo{volume}{B667}},
  \bibinfo{pages}{201} (\bibinfo{year}{2003}).

\bibitem[{\citenamefont{Ji et~al.}(2004)\citenamefont{Ji, Ma, and
  Yuan}}]{TMDFactJi}
\bibinfo{author}{\bibfnamefont{X.-d.} \bibnamefont{Ji}},
  \bibinfo{author}{\bibfnamefont{J.-P.} \bibnamefont{Ma}}, \bibnamefont{and}
  \bibinfo{author}{\bibfnamefont{F.}~\bibnamefont{Yuan}},
  \bibinfo{journal}{Phys. Lett. B} \textbf{\bibinfo{volume}{597}},
  \bibinfo{pages}{299} (\bibinfo{year}{2004});
\bibitem[{\citenamefont{Ji et~al.}(2005)\citenamefont{Ji, Ma, and
  Yuan}}]{TMDFactJi2}
  \bibinfo{journal}{Phys. Rev. D} \textbf{\bibinfo{volume}{71}},
  \bibinfo{pages}{034005} (\bibinfo{year}{2005}).

\bibitem[{\citenamefont{Collins and Metz}(2004)}]{TMDFactCollinsMetz}
\bibinfo{author}{\bibfnamefont{J.~C.} \bibnamefont{Collins}} \bibnamefont{and}
  \bibinfo{author}{\bibfnamefont{A.}~\bibnamefont{Metz}},
  \bibinfo{journal}{Phys. Rev. Lett.} \textbf{\bibinfo{volume}{93}},
  \bibinfo{pages}{252001} (\bibinfo{year}{2004}).

\bibitem[{\citenamefont{Kang and Qiu}(2009)}]{WANKang}
\bibinfo{author}{\bibfnamefont{Z.-B.} \bibnamefont{Kang}} \bibnamefont{and}
  \bibinfo{author}{\bibfnamefont{J.-W.} \bibnamefont{Qiu}},
  \bibinfo{journal}{Phys. Rev. Lett.} \textbf{\bibinfo{volume}{103}},
  \bibinfo{pages}{172001} (\bibinfo{year}{2009}).

\bibitem[{\citenamefont{Collins and Qiu}(2007)}]{TMDFactBreakCollinsQiu}
\bibinfo{author}{\bibfnamefont{J.}~\bibnamefont{Collins}} \bibnamefont{and}
  \bibinfo{author}{\bibfnamefont{J.-W.} \bibnamefont{Qiu}},
  \bibinfo{journal}{Phys. Rev. D} \textbf{\bibinfo{volume}{75}},
  \bibinfo{pages}{114014} (\bibinfo{year}{2007});
\bibitem[{\citenamefont{Collins}(2007)}]{TMDFactBreakCollins}
\bibinfo{author}{\bibfnamefont{J.}~\bibnamefont{Collins}}, \eprint{arXiv:0708.4410}.

\bibitem[{\citenamefont{Rogers and Mulders}(2010)}]{TMDFactBreakRogers}
\bibinfo{author}{\bibfnamefont{T.~C.} \bibnamefont{Rogers}} \bibnamefont{and}
  \bibinfo{author}{\bibfnamefont{P.~J.} \bibnamefont{Mulders}},
  \bibinfo{journal}{Phys. Rev. D} \textbf{\bibinfo{volume}{81}},
  \bibinfo{pages}{094006} (\bibinfo{year}{2010}).

\bibitem[{\citenamefont{Kang et~al.}(2017{\natexlab{a}})\citenamefont{Kang,
  Liu, Ringer, and Xing}}]{TMDFF}
\bibinfo{author}{\bibfnamefont{Z.-B.} \bibnamefont{Kang}},
  \bibinfo{author}{\bibfnamefont{X.}~\bibnamefont{Liu}},
  \bibinfo{author}{\bibfnamefont{F.}~\bibnamefont{Ringer}}, \bibnamefont{and}
  \bibinfo{author}{\bibfnamefont{H.}~\bibnamefont{Xing}}, \eprint{arXiv:1705.08443}.

\bibitem[{\citenamefont{Ji et~al.}(2006{\natexlab{a}})\citenamefont{Ji, Qiu,
  Vogelsang, and Yuan}}]{Twist3TMDConsistent1}
\bibinfo{author}{\bibfnamefont{X.}~\bibnamefont{Ji}},
  \bibinfo{author}{\bibfnamefont{J.~W.} \bibnamefont{Qiu}},
  \bibinfo{author}{\bibfnamefont{W.}~\bibnamefont{Vogelsang}},
  \bibnamefont{and} \bibinfo{author}{\bibfnamefont{F.}~\bibnamefont{Yuan}},
  \bibinfo{journal}{Phys. Rev. Lett.} \textbf{\bibinfo{volume}{97}},
  \bibinfo{pages}{082002} (\bibinfo{year}{2006}{\natexlab{a}});
\bibitem[{\citenamefont{Ji et~al.}(2006{\natexlab{b}})\citenamefont{Ji, Qiu,
  Vogelsang, and Yuan}}]{Twist3TMDConsistent2}
  \bibinfo{journal}{Phys. Rev. D} \textbf{\bibinfo{volume}{73}},
  \bibinfo{pages}{094017} (\bibinfo{year}{2006}{\natexlab{b}});
\bibitem[{\citenamefont{Ji et~al.}(2006{\natexlab{c}})\citenamefont{Ji, Qiu,
  Vogelsang, and Yuan}}]{Twist3TMDConsistent3}
  \bibinfo{journal}{Phys. Lett. B} \textbf{\bibinfo{volume}{638}},
  \bibinfo{pages}{178} (\bibinfo{year}{2006}{\natexlab{c}}).

\bibitem[{\citenamefont{Gamberg et~al.}(2013)\citenamefont{Gamberg, Kang, and
  Prokudin}}]{GambergSivers}
\bibinfo{author}{\bibfnamefont{L.}~\bibnamefont{Gamberg}},
  \bibinfo{author}{\bibfnamefont{Z.-B.} \bibnamefont{Kang}}, \bibnamefont{and}
  \bibinfo{author}{\bibfnamefont{A.}~\bibnamefont{Prokudin}},
  \bibinfo{journal}{Phys. Rev. Lett.} \textbf{\bibinfo{volume}{110}},
  \bibinfo{pages}{232301} (\bibinfo{year}{2013}).

\bibitem[{\citenamefont{Airapetian et~al.}(2005)}]{HERMESColSiv}
\bibinfo{author}{\bibfnamefont{A.}~\bibnamefont{Airapetian}}
  \bibnamefont{et~al.} (\bibinfo{collaboration}{HERMES Collaboration}),
  \bibinfo{journal}{Phys. Rev. Lett.} \textbf{\bibinfo{volume}{94}},
  \bibinfo{pages}{012002} (\bibinfo{year}{2005});
\bibitem[{\citenamefont{Airapetian et~al.}(2010)}]{HERMESCol10}
  \bibinfo{journal}{Phys. Lett. B} \textbf{\bibinfo{volume}{693}},
  \bibinfo{pages}{11} (\bibinfo{year}{2010}).

\bibitem[{\citenamefont{Airapetian et~al.}(2009)}]{HERMESSiv09}
\bibinfo{author}{\bibfnamefont{A.}~\bibnamefont{Airapetian}}
  \bibnamefont{et~al.} (\bibinfo{collaboration}{HERMES Collaboration}),
  \bibinfo{journal}{Phys. Rev. Lett.} \textbf{\bibinfo{volume}{103}},
  \bibinfo{pages}{152002} (\bibinfo{year}{2009}).

\bibitem[{\citenamefont{Alexakhin et~al.}(2005)}]{COMPASSColSiv05}
\bibinfo{author}{\bibfnamefont{V.~Y.} \bibnamefont{Alexakhin}}
  \bibnamefont{et~al.} (\bibinfo{collaboration}{COMPASS Collaboration}),
  \bibinfo{journal}{Phys. Rev. Lett.} \textbf{\bibinfo{volume}{94}},
  \bibinfo{pages}{202002} (\bibinfo{year}{2005});
  \bibitem[{\citenamefont{Ageev et~al.}(2007)}]{COMPASSColSiv07}
\bibinfo{author}{\bibfnamefont{E.~S.} \bibnamefont{Ageev}} \bibnamefont{et~al.}
  (\bibinfo{collaboration}{COMPASS Collaboration}), \bibinfo{journal}{Nucl.
  Phys.} \textbf{\bibinfo{volume}{B765}}, \bibinfo{pages}{31}
  (\bibinfo{year}{2007});
\bibitem[{\citenamefont{Alekseev et~al.}(2009)}]{COMPASSColSiv09}
\bibinfo{author}{\bibfnamefont{M.~G.} \bibnamefont{Alekseev}}
  \bibnamefont{et~al.} (\bibinfo{collaboration}{COMPASS Collaboration}),
  \bibinfo{journal}{Phys. Lett. B} \textbf{\bibinfo{volume}{673}},
  \bibinfo{pages}{127} (\bibinfo{year}{2009});
\bibitem[{\citenamefont{Alekseev et~al.}(2010)}]{COMPASSColSiv10}
 \textbf{\bibinfo{volume}{692}},
  \bibinfo{pages}{240} (\bibinfo{year}{2010});
\bibitem[{\citenamefont{Adolph et~al.}(2012)}]{COMPASSCol12}
\bibinfo{author}{\bibfnamefont{C.}~\bibnamefont{Adolph}} \bibnamefont{et~al.}
  (\bibinfo{collaboration}{COMPASS Collaboration}), \bibinfo{journal}{\textit{ibid.}} \textbf{\bibinfo{volume}{717}}, \bibinfo{pages}{376}
  (\bibinfo{year}{2012});
\bibitem[{\citenamefont{Adolph et~al.}(2015)}]{COMPASSnew}
 \textbf{\bibinfo{volume}{744}}, \bibinfo{pages}{250}
  (\bibinfo{year}{2015}).

\bibitem[{\citenamefont{Qian et~al.}(2011)}]{JLABColSiv}
\bibinfo{author}{\bibfnamefont{X.}~\bibnamefont{Qian}} \bibnamefont{et~al.}
  (\bibinfo{collaboration}{Jefferson Lab Hall A Collaboration}),
  \bibinfo{journal}{Phys. Rev. Lett.} \textbf{\bibinfo{volume}{107}},
  \bibinfo{pages}{072003} (\bibinfo{year}{2011}).

\bibitem[{\citenamefont{Seidl et~al.}(2006)}]{BELLE2006}
\bibinfo{author}{\bibfnamefont{R.}~\bibnamefont{Seidl}} \bibnamefont{et~al.}
  (\bibinfo{collaboration}{Belle Collaboration}), \bibinfo{journal}{Phys. Rev.
  Lett.} \textbf{\bibinfo{volume}{96}}, \bibinfo{pages}{232002}
  (\bibinfo{year}{2006});
\bibitem[{\citenamefont{Seidl et~al.}(2012)}]{BELLE2012}
\bibinfo{journal}{Phys. Rev.
  D} \textbf{\bibinfo{volume}{86}}, \bibinfo{pages}{039905(E)}
  (\bibinfo{year}{2012}).

\bibitem[{\citenamefont{Lees et~al.}(2014)}]{BABAR}
\bibinfo{author}{\bibfnamefont{J.~P.} \bibnamefont{Lees}} \bibnamefont{et~al.}
  (\bibinfo{collaboration}{BaBar Collaboration}), \bibinfo{journal}{Phys. Rev.
  D} \textbf{\bibinfo{volume}{90}}, \bibinfo{pages}{052003}
  (\bibinfo{year}{2014}).

\bibitem[{\citenamefont{Anselmino et~al.}(2007)\citenamefont{Anselmino,
  Boglione, D'Alesio, Kotzinian, Murgia, and Prokudin}}]{Transversity07}
\bibinfo{author}{\bibfnamefont{M.}~\bibnamefont{Anselmino}},
  \bibinfo{author}{\bibfnamefont{M.}~\bibnamefont{Boglione}},
  \bibinfo{author}{\bibfnamefont{U.}~\bibnamefont{D'Alesio}},
  \bibinfo{author}{\bibfnamefont{A.}~\bibnamefont{Kotzinian}},
  \bibinfo{author}{\bibfnamefont{F.}~\bibnamefont{Murgia}},
  \bibinfo{author}{\bibfnamefont{A.}~\bibnamefont{Prokudin}}, \bibnamefont{and}
  \bibinfo{author}{\bibfnamefont{C.}~\bibnamefont{Turk}},
  \bibinfo{journal}{Phys. Rev. D} \textbf{\bibinfo{volume}{75}},
  \bibinfo{pages}{054032} (\bibinfo{year}{2007}).

\bibitem[{\citenamefont{Anselmino et~al.}(2009)\citenamefont{Anselmino,
  Boglione, D'Alesio, Kotzinian, Melis, Murgia, and Prokudin}}]{Transversity09}
\bibinfo{author}{\bibfnamefont{M.}~\bibnamefont{Anselmino}},
  \bibinfo{author}{\bibfnamefont{M.}~\bibnamefont{Boglione}},
  \bibinfo{author}{\bibfnamefont{U.}~\bibnamefont{D'Alesio}},
  \bibinfo{author}{\bibfnamefont{F.}~\bibnamefont{Kotzinian}},
  \bibinfo{author}{\bibfnamefont{S.}~\bibnamefont{Melis}},
  \bibinfo{author}{\bibfnamefont{F.}~\bibnamefont{Murgia}}, \bibnamefont{and}
  \bibinfo{author}{\bibfnamefont{A.}~\bibnamefont{Prokudin}},
  \bibinfo{journal}{Nucl.~Phys.~B, Proc.~Suppl.}
  \textbf{\bibinfo{volume}{191}}, \bibinfo{pages}{98} (\bibinfo{year}{2009}).

\bibitem[{\citenamefont{Anselmino et~al.}(2013)\citenamefont{Anselmino,
  Boglione, D'Alesio, Melis, Murgia, and Prokudin}}]{Transversity13}
\bibinfo{author}{\bibfnamefont{M.}~\bibnamefont{Anselmino}},
  \bibinfo{author}{\bibfnamefont{M.}~\bibnamefont{Boglione}},
  \bibinfo{author}{\bibfnamefont{U.}~\bibnamefont{D'Alesio}},
  \bibinfo{author}{\bibfnamefont{S.}~\bibnamefont{Melis}},
  \bibinfo{author}{\bibfnamefont{F.}~\bibnamefont{Murgia}}, \bibnamefont{and}
  \bibinfo{author}{\bibfnamefont{A.}~\bibnamefont{Prokudin}},
  \bibinfo{journal}{Phys. Rev. D} \textbf{\bibinfo{volume}{87}},
  \bibinfo{pages}{094019} (\bibinfo{year}{2013}).

\bibitem[{\citenamefont{Anselmino et~al.}(2015)\citenamefont{Anselmino,
  Boglione, D'Alesio, Hernandez, Melis, Murgia, and Prokudin}}]{Transversity15}
\bibinfo{author}{\bibfnamefont{M.}~\bibnamefont{Anselmino}},
  \bibinfo{author}{\bibfnamefont{M.}~\bibnamefont{Boglione}},
  \bibinfo{author}{\bibfnamefont{U.}~\bibnamefont{D'Alesio}},
  \bibinfo{author}{\bibfnamefont{J.~O.~Gonzalez} \bibnamefont{Hernandez}},
  \bibinfo{author}{\bibfnamefont{S.}~\bibnamefont{Melis}},
  \bibinfo{author}{\bibfnamefont{F.}~\bibnamefont{Murgia}}, \bibnamefont{and}
  \bibinfo{author}{\bibfnamefont{A.}~\bibnamefont{Prokudin}},
  \bibinfo{journal}{Phys. Rev. D} \textbf{\bibinfo{volume}{92}},
  \bibinfo{pages}{114023} (\bibinfo{year}{2015}).

\bibitem[{\citenamefont{Kang et~al.}(2016)\citenamefont{Kang, Prokudin, Sun,
  and Yuan}}]{TransversityKang}
\bibinfo{author}{\bibfnamefont{Z.-B.} \bibnamefont{Kang}},
  \bibinfo{author}{\bibfnamefont{A.}~\bibnamefont{Prokudin}},
  \bibinfo{author}{\bibfnamefont{P.}~\bibnamefont{Sun}}, \bibnamefont{and}
  \bibinfo{author}{\bibfnamefont{F.}~\bibnamefont{Yuan}},
  \bibinfo{journal}{Phys. Rev. D} \textbf{\bibinfo{volume}{93}},
  \bibinfo{pages}{014009} (\bibinfo{year}{2016}).

\bibitem[{\citenamefont{D'Alesio et~al.}(2011)\citenamefont{D'Alesio, Murgia,
  and Pisano}}]{D'Alesio_2011}
\bibinfo{author}{\bibfnamefont{U.}~\bibnamefont{D'Alesio}},
  \bibinfo{author}{\bibfnamefont{F.}~\bibnamefont{Murgia}}, \bibnamefont{and}
  \bibinfo{author}{\bibfnamefont{C.}~\bibnamefont{Pisano}},
  \bibinfo{journal}{Phys. Rev. D} \textbf{\bibinfo{volume}{83}},
  \bibinfo{pages}{034021} (\bibinfo{year}{2011}).

\bibitem[{\citenamefont{Anselmino et~al.}(2006)}]{GPM_2006}
\bibinfo{author}{\bibfnamefont{M.}~\bibnamefont{Anselmino}},
\bibinfo{author}{\bibfnamefont{M.}~\bibnamefont{Boglione}},
\bibinfo{author}{\bibfnamefont{U.}~\bibnamefont{D'Alesio}},
\bibinfo{author}{\bibfnamefont{E.}~\bibnamefont{Leader}},
\bibinfo{author}{\bibfnamefont{S.}~\bibnamefont{Melis}}, \bibnamefont{and}
\bibinfo{author}{\bibfnamefont{F.}~\bibnamefont{Murgia}},
  \bibinfo{journal}{Phys. Rev. D}
  \textbf{\bibinfo{volume}{73}}, \bibinfo{pages}{014020}
  (\bibinfo{year}{2006}).

\bibitem[{\citenamefont{Yuan}(2008{\natexlab{a}})}]{YuanCollins}
\bibinfo{author}{\bibfnamefont{F.}~\bibnamefont{Yuan}}, \bibinfo{journal}{Phys.
  Rev. Lett.} \textbf{\bibinfo{volume}{100}}, \bibinfo{pages}{032003}
  (\bibinfo{year}{2008}{\natexlab{a}});
\bibitem[{\citenamefont{Yuan}(2008{\natexlab{b}})}]{YuanCollinsLong}
\bibinfo{journal}{Phys.
  Rev. D} \textbf{\bibinfo{volume}{77}}, \bibinfo{pages}{074019}
  (\bibinfo{year}{2008}{\natexlab{b}}).

\bibitem[{\citenamefont{Boer and Mulders}(1998)}]{BoerMulders}
\bibinfo{author}{\bibfnamefont{D.}~\bibnamefont{Boer}} \bibnamefont{and}
  \bibinfo{author}{\bibfnamefont{P.~J.} \bibnamefont{Mulders}},
  \bibinfo{journal}{Phys. Rev. D} \textbf{\bibinfo{volume}{57}},
  \bibinfo{pages}{5780} (\bibinfo{year}{1998}).

\bibitem[{\citenamefont{Mukherjee and Vogelsang}(2012)}]{NLO}
\bibinfo{author}{\bibfnamefont{A.}~\bibnamefont{Mukherjee}} \bibnamefont{and}
  \bibinfo{author}{\bibfnamefont{W.}~\bibnamefont{Vogelsang}},
  \bibinfo{journal}{Phys. Rev. D} \textbf{\bibinfo{volume}{86}},
  \bibinfo{pages}{094009} (\bibinfo{year}{2012}).

\bibitem[{\citenamefont{Lai et~al.}(2010)\citenamefont{Lai, Guzzi, Huston, Li,
  Nadolsky, Pumplin, and Yuan}}]{CT10}
\bibinfo{author}{\bibfnamefont{H.-L.} \bibnamefont{Lai}},
  \bibinfo{author}{\bibfnamefont{M.}~\bibnamefont{Guzzi}},
  \bibinfo{author}{\bibfnamefont{J.}~\bibnamefont{Huston}},
  \bibinfo{author}{\bibfnamefont{Z.}~\bibnamefont{Li}},
  \bibinfo{author}{\bibfnamefont{P.~M.} \bibnamefont{Nadolsky}},
  \bibinfo{author}{\bibfnamefont{J.}~\bibnamefont{Pumplin}}, \bibnamefont{and}
  \bibinfo{author}{\bibfnamefont{C.-P.} \bibnamefont{Yuan}},
  \bibinfo{journal}{Phys. Rev. D} \textbf{\bibinfo{volume}{82}},
  \bibinfo{pages}{074024} (\bibinfo{year}{2010}).

\bibitem[{\citenamefont{Boer et~al.}(2015)\citenamefont{Boer, Lorc\'{e},
  Pisano, and Zhou}}]{BoerGluonSivers}
\bibinfo{author}{\bibfnamefont{D.}~\bibnamefont{Boer}},
  \bibinfo{author}{\bibfnamefont{C.}~\bibnamefont{Lorc\'{e}}},
  \bibinfo{author}{\bibfnamefont{C.}~\bibnamefont{Pisano}}, \bibnamefont{and}
  \bibinfo{author}{\bibfnamefont{J.}~\bibnamefont{Zhou}},
  \bibinfo{journal}{Adv.~High Energy Phys.} \textbf{\bibinfo{volume}{2015}},
  \bibinfo{pages}{371396} (\bibinfo{year}{2015}).

\bibitem[{\citenamefont{Adamczyk et~al.}(2012{\natexlab{b}})}]{STAR_jet_AN2}
\bibinfo{author}{\bibfnamefont{L.}~\bibnamefont{Adamczyk}} \bibnamefont{et~al.}
  (\bibinfo{collaboration}{STAR Collaboration}), \bibinfo{journal}{Phys. Rev.
  D} \textbf{\bibinfo{volume}{86}}, \bibinfo{pages}{032006}
  (\bibinfo{year}{2012}{\natexlab{b}}).

\bibitem[{\citenamefont{Harrison et~al.}(2003)\citenamefont{Harrison, Ludlam,
  and Ozaki}}]{RHIC_NIM}
\bibinfo{author}{\bibfnamefont{M.}~\bibnamefont{Harrison}},
  \bibinfo{author}{\bibfnamefont{T.}~\bibnamefont{Ludlam}}, \bibnamefont{and}
  \bibinfo{author}{\bibfnamefont{S.}~\bibnamefont{Ozaki}},
  \bibinfo{journal}{Nucl. Inst. \& Meth.} \textbf{\bibinfo{volume}{A499}},
  \bibinfo{pages}{235} (\bibinfo{year}{2003});
\bibitem[{\citenamefont{Hahn et~al.}(2003)}]{RHIC_NIM2}
\bibinfo{author}{\bibfnamefont{H.}~\bibnamefont{Hahn}} \bibnamefont{et~al.},
  \bibinfo{journal}{\textit{ibid.}} \textbf{\bibinfo{volume}{A499}},
  \bibinfo{pages}{245} (\bibinfo{year}{2003}).

\bibitem[{\citenamefont{Ackermann et~al.}(2003)}]{STAR_NIM}
\bibinfo{author}{\bibfnamefont{K.~H.} \bibnamefont{Ackermann}}
  \bibnamefont{et~al.} (\bibinfo{collaboration}{STAR Collaboration}),
  \bibinfo{journal}{Nucl. Inst. \& Meth.} \textbf{\bibinfo{volume}{A499}},
  \bibinfo{pages}{624} (\bibinfo{year}{2003}).

\bibitem[{\citenamefont{Alekseev et~al.}(2003)}]{RHIC-pp_NIM}
\bibinfo{author}{\bibfnamefont{I.}~\bibnamefont{Alekseev}}
  \bibnamefont{et~al.}, \bibinfo{journal}{Nucl. Inst. \& Meth.}
  \textbf{\bibinfo{volume}{A499}}, \bibinfo{pages}{392} (\bibinfo{year}{2003}).

\bibitem[{\citenamefont{\mbox{RHIC Polarimetry} Group}(2013)}]{2011Pol}
\bibinfo{author}{\bibnamefont{\mbox{RHIC Polarimetry} Group}},
  \emph{\bibinfo{title}{\mbox{RHIC} \mbox{P}olarizations for \mbox{R}uns 9-12}}
  (\bibinfo{year}{2013}), \bibinfo{note}{\mbox{C}-A/AP/490}.

\bibitem[{\citenamefont{Jinnouchi et~al.}(2005)}]{pC_Pol}
\bibinfo{author}{\bibfnamefont{O.}~\bibnamefont{Jinnouchi}}
  \bibnamefont{et~al.}, in \emph{\bibinfo{booktitle}{Proceedings of the 16th
  International Symposium on Spin Physics (Spin2004)}} (\bibinfo{year}{2005}).

\bibitem[{\citenamefont{Okada et~al.}(2006)}]{HJet_Pol}
\bibinfo{author}{\bibfnamefont{H.}~\bibnamefont{Okada}} \bibnamefont{et~al.},
  \bibinfo{journal}{Phys. Lett. B} \textbf{\bibinfo{volume}{638}},
  \bibinfo{pages}{450} (\bibinfo{year}{2006}).

\bibitem[{\citenamefont{Bieser et~al.}(2003)}]{STAR_TRG}
\bibinfo{author}{\bibfnamefont{F.~S.} \bibnamefont{Bieser}}
  \bibnamefont{et~al.}, \bibinfo{journal}{Nucl. Inst. \& Meth.}
  \textbf{\bibinfo{volume}{A499}}, \bibinfo{pages}{766} (\bibinfo{year}{2003}).

\bibitem[{\citenamefont{Llope et~al.}(2014)}]{VPD_NIM}
\bibinfo{author}{\bibfnamefont{W.~J.} \bibnamefont{Llope}}
  \bibnamefont{et~al.}, \bibinfo{journal}{Nucl. Inst. \& Meth.}
  \textbf{\bibinfo{volume}{A759}}, \bibinfo{pages}{23} (\bibinfo{year}{2014}).

\bibitem[{\citenamefont{Anderson et~al.}(2003)}]{TPC_NIM}
\bibinfo{author}{\bibfnamefont{W.}~\bibnamefont{Anderson}}
  \bibnamefont{et~al.}, \bibinfo{journal}{Nucl. Inst. \& Meth.}
  \textbf{\bibinfo{volume}{A499}}, \bibinfo{pages}{659} (\bibinfo{year}{2003}).

\bibitem[{\citenamefont{Llope}(2005)}]{TOF_NIM}
\bibinfo{author}{\bibfnamefont{W.~J.} \bibnamefont{Llope}},
  \bibinfo{journal}{Nucl. Inst. \& Meth.} \textbf{\bibinfo{volume}{B241}},
  \bibinfo{pages}{306} (\bibinfo{year}{2005}).

\bibitem[{\citenamefont{Shao et~al.}(2006)}]{PID}
\bibinfo{author}{\bibfnamefont{M.}~\bibnamefont{Shao}} \bibnamefont{et~al.},
  \bibinfo{journal}{Nucl. Inst. \& Meth.} \textbf{\bibinfo{volume}{A558}},
  \bibinfo{pages}{419} (\bibinfo{year}{2006});
\bibitem[{\citenamefont{Xu et~al.}(2010)}]{PID_highPt}
\bibinfo{author}{\bibfnamefont{Y.}~\bibnamefont{Xu}} \bibnamefont{et~al.},
  \bibinfo{journal}{\textit{ibid.}} \textbf{\bibinfo{volume}{A614}},
  \bibinfo{pages}{28} (\bibinfo{year}{2010}).

\bibitem[{\citenamefont{Beddo et~al.}(2003)}]{BEMC_NIM}
\bibinfo{author}{\bibfnamefont{M.}~\bibnamefont{Beddo}} \bibnamefont{et~al.},
  \bibinfo{journal}{Nucl. Inst. \& Meth.} \textbf{\bibinfo{volume}{A499}},
  \bibinfo{pages}{725} (\bibinfo{year}{2003}).

\bibitem[{\citenamefont{Allgower et~al.}(2003)}]{EEMC_NIM}
\bibinfo{author}{\bibfnamefont{C.~E.} \bibnamefont{Allgower}}
  \bibnamefont{et~al.}, \bibinfo{journal}{Nucl. Inst. \& Meth.}
  \textbf{\bibinfo{volume}{A499}}, \bibinfo{pages}{740} (\bibinfo{year}{2003}).

\bibitem[{\citenamefont{Margetis and Cebra}(1994)}]{STARVtxNote}
\bibinfo{author}{\bibfnamefont{S.}~\bibnamefont{Margetis}} \bibnamefont{and}
  \bibinfo{author}{\bibfnamefont{D.}~\bibnamefont{Cebra}},
  \bibinfo{journal}{STARNote SN0089}  (\bibinfo{year}{1994});
\bibitem[{\citenamefont{Reed et~al.}(2010)}]{STARVtx}
\bibinfo{author}{\bibfnamefont{R.}~\bibnamefont{Reed}} \bibnamefont{et~al.},
  \bibinfo{journal}{J.~Phys.~Conf.~Ser.} \textbf{\bibinfo{volume}{219}},
  \bibinfo{pages}{032020} (\bibinfo{year}{2010}).

\bibitem[{\citenamefont{Adamczyk et~al.}(2015{\natexlab{a}})}]{2009ALL}
\bibinfo{author}{\bibfnamefont{L.}~\bibnamefont{Adamczyk}} \bibnamefont{et~al.}
  (\bibinfo{collaboration}{STAR Collaboration}), \bibinfo{journal}{Phys. Rev.
  Lett.} \textbf{\bibinfo{volume}{115}}, \bibinfo{pages}{092002}
  (\bibinfo{year}{2015}{\natexlab{a}}).

\bibitem[{\citenamefont{Cacciari et~al.}(2008)\citenamefont{Cacciari, Salam,
  and Soyez}}]{AntiKt}
\bibinfo{author}{\bibfnamefont{M.}~\bibnamefont{Cacciari}},
  \bibinfo{author}{\bibfnamefont{G.~P.} \bibnamefont{Salam}}, \bibnamefont{and}
  \bibinfo{author}{\bibfnamefont{G.}~\bibnamefont{Soyez}},
  \bibinfo{journal}{J.~High Energy Phys.} \textbf{\bibinfo{volume}{04}},
  \bibinfo{pages}{063} (\bibinfo{year}{2008}).

\bibitem[{\citenamefont{Cacciari et~al.}(2012)\citenamefont{Cacciari, Salam,
  and Soyez}}]{FastJet}
\bibinfo{author}{\bibfnamefont{M.}~\bibnamefont{Cacciari}},
  \bibinfo{author}{\bibfnamefont{G.~P.} \bibnamefont{Salam}}, \bibnamefont{and}
  \bibinfo{author}{\bibfnamefont{G.}~\bibnamefont{Soyez}},
  \bibinfo{journal}{Eur.~Phys.~J.} \textbf{\bibinfo{volume}{C72}},
  \bibinfo{pages}{1896} (\bibinfo{year}{2012}).

\bibitem[{\citenamefont{Ohlsen and Keaton}(1973)}]{CrossRatio}
\bibinfo{author}{\bibfnamefont{G.~G.} \bibnamefont{Ohlsen}} \bibnamefont{and}
  \bibinfo{author}{\bibfnamefont{P.~W.} \bibnamefont{Keaton}},
  \bibinfo{journal}{Nucl. Inst. \& Meth.} \textbf{\bibinfo{volume}{109}},
  \bibinfo{pages}{41} (\bibinfo{year}{1973}).

\bibitem[{\citenamefont{Sj\"{o}strand et~al.}(2006)\citenamefont{Sj\"{o}strand,
  Mrenna, and Skands}}]{Pythia}
\bibinfo{author}{\bibfnamefont{T.}~\bibnamefont{Sj\"{o}strand}},
  \bibinfo{author}{\bibfnamefont{S.}~\bibnamefont{Mrenna}}, \bibnamefont{and}
  \bibinfo{author}{\bibfnamefont{P.}~\bibnamefont{Skands}},
  \bibinfo{journal}{J. High Energy Phys.} \textbf{\bibinfo{volume}{05}},
  \bibinfo{pages}{026} (\bibinfo{year}{2006}).

\bibitem[{\citenamefont{Skands}(2010)}]{Pythia2}
\bibinfo{author}{\bibfnamefont{P.~Z.} \bibnamefont{Skands}},
  \bibinfo{journal}{Phys. Rev. D} \textbf{\bibinfo{volume}{82}},
  \bibinfo{pages}{074018} (\bibinfo{year}{2010}).

\bibitem[{\citenamefont{Brun et~al.}(1987)}]{GEANT0}
\bibinfo{author}{\bibfnamefont{R.}~\bibnamefont{Brun}} \bibnamefont{et~al.},
  \emph{\bibinfo{title}{\mbox{GEANT3}}} (\bibinfo{year}{1987}),
  \bibinfo{note}{\mbox{CERN-DD-EE-84-1}};
\bibitem[{\citenamefont{Brun et~al.}(1994)\citenamefont{Brun, Carminati, and
  Giani}}]{GEANT1}
\bibinfo{author}{\bibfnamefont{R.}~\bibnamefont{Brun}},
  \bibinfo{author}{\bibfnamefont{F.}~\bibnamefont{Carminati}},
  \bibnamefont{and} \bibinfo{author}{\bibfnamefont{S.}~\bibnamefont{Giani}},
  \emph{\bibinfo{title}{\mbox{GEANT} \mbox{Detector} \mbox{Description} and
  \mbox{Simulation} \mbox{Tool}}} (\bibinfo{year}{1994}),
  \bibinfo{note}{\mbox{CERN-W5013}, \mbox{CERN-W-5013}}.

\bibitem[{\citenamefont{Leader}(2011)}]{LeaderSpinBook}
\bibinfo{author}{\bibfnamefont{E.}~\bibnamefont{Leader}},
  \bibinfo{journal}{Camb.~Monogr.~Part.~Phys.~Nucl.~Phys.~Cosmol.}
  \textbf{\bibinfo{volume}{15}} (\bibinfo{year}{2011}).

\bibitem[{\citenamefont{\mbox{STAR} \mbox{ZCD}~\mbox{Polarimetry}
  Group}()}]{RadialPol}
\bibinfo{author}{\bibnamefont{\mbox{STAR} \mbox{ZCD}~\mbox{Polarimetry}
  Group}}, \emph{\bibinfo{title}{\mbox{ZDC} \mbox{Scaler} \mbox{Polarimetry}}},
  \bibinfo{note}{\\http://online.star.bnl.gov/scaler2011/polarimetry/}.

\bibitem[{\citenamefont{D'Alesio et~al.}(2017)\citenamefont{D'Alesio, Murgia,
  and Pisano}}]{D'Alesio_2017}
\bibinfo{author}{\bibfnamefont{U.}~\bibnamefont{D'Alesio}},
  \bibinfo{author}{\bibfnamefont{F.}~\bibnamefont{Murgia}}, \bibnamefont{and}
  \bibinfo{author}{\bibfnamefont{C.}~\bibnamefont{Pisano}},
  \bibinfo{journal}{Phys. Lett. B} \textbf{\bibinfo{volume}{773}},
  \bibinfo{pages}{300} (\bibinfo{year}{2017}).

\bibitem[{\citenamefont{Kretzer}(2000)}]{KretzerFF}
\bibinfo{author}{\bibfnamefont{S.}~\bibnamefont{Kretzer}},
  \bibinfo{journal}{Phys. Rev. D} \textbf{\bibinfo{volume}{62}},
  \bibinfo{pages}{054001} (\bibinfo{year}{2000}).

\bibitem[{\citenamefont{de~Florian et~al.}(2007)\citenamefont{de~Florian,
  Sassot, and Stratmann}}]{DSS}
\bibinfo{author}{\bibfnamefont{D.}~\bibnamefont{de~Florian}},
  \bibinfo{author}{\bibfnamefont{R.}~\bibnamefont{Sassot}}, \bibnamefont{and}
  \bibinfo{author}{\bibfnamefont{M.}~\bibnamefont{Stratmann}},
  \bibinfo{journal}{Phys. Rev. D} \textbf{\bibinfo{volume}{75}},
  \bibinfo{pages}{114010} (\bibinfo{year}{2007}).

\bibitem[{\citenamefont{Kang et~al.}(2017{\natexlab{b}})\citenamefont{Kang,
  Prokudin, Ringer, and Yuan}}]{KPRY}
\bibinfo{author}{\bibfnamefont{Z.-B.} \bibnamefont{Kang}},
  \bibinfo{author}{\bibfnamefont{A.}~\bibnamefont{Prokudin}},
  \bibinfo{author}{\bibfnamefont{F.}~\bibnamefont{Ringer}}, \bibnamefont{and}
  \bibinfo{author}{\bibfnamefont{F.}~\bibnamefont{Yuan}},
  \bibinfo{journal}{Phys. Lett. B} \textbf{\bibinfo{volume}{774}},
  \bibinfo{pages}{635} (\bibinfo{year}{2017}{\natexlab{b}}).

\bibitem[{\citenamefont{de~Florian et~al.}(2015)\citenamefont{de~Florian,
  Sassot, Epele, Hern{\'a}ndez-Pinto, and Stratmann}}]{DSS14}
\bibinfo{author}{\bibfnamefont{D.}~\bibnamefont{de~Florian}},
  \bibinfo{author}{\bibfnamefont{R.}~\bibnamefont{Sassot}},
  \bibinfo{author}{\bibfnamefont{M.}~\bibnamefont{Epele}},
  \bibinfo{author}{\bibfnamefont{R.~J.} \bibnamefont{Hern{\'a}ndez-Pinto}},
  \bibnamefont{and}
  \bibinfo{author}{\bibfnamefont{M.}~\bibnamefont{Stratmann}},
  \bibinfo{journal}{Phys. Rev. D} \textbf{\bibinfo{volume}{91}},
  \bibinfo{pages}{014035} (\bibinfo{year}{2015}).

\bibitem[{\citenamefont{Adamczyk et~al.}(2015{\natexlab{b}})}]{STAR_IFF2006}
\bibinfo{author}{\bibfnamefont{L.}~\bibnamefont{Adamczyk}} \bibnamefont{et~al.}
  (\bibinfo{collaboration}{STAR Collaboration}), \bibinfo{journal}{Phys. Rev.
  Lett.} \textbf{\bibinfo{volume}{115}}, \bibinfo{pages}{242501}
  (\bibinfo{year}{2015}{\natexlab{b}});
\bibitem[{\citenamefont{Adams et~al.}(2017)}]{STAR_IFF2011}
\bibinfo{author}{\bibfnamefont{J.}~\bibnamefont{Adams}} \bibnamefont{et~al.}
  (\bibinfo{collaboration}{STAR Collaboration}),
  \eprint{arXiv:1710.10215}.

\bibitem[{\citenamefont{Bacchetta and Radici}(2004)}]{BacchettaRadici}
\bibinfo{author}{\bibfnamefont{A.}~\bibnamefont{Bacchetta}} \bibnamefont{and}
  \bibinfo{author}{\bibfnamefont{M.}~\bibnamefont{Radici}},
  \bibinfo{journal}{Phys. Rev. D} \textbf{\bibinfo{volume}{70}},
  \bibinfo{pages}{094032} (\bibinfo{year}{2004}).

\bibitem[{\citenamefont{Radici}(2017)}]{RadiciDIS2017}
\bibinfo{author}{\bibfnamefont{M.}~\bibnamefont{Radici}}, in
  \emph{\bibinfo{booktitle}{25th International Workshop on Deep Inelastic
  Scattering and Related Topics (DIS 2017) Birmingham, UK, April 3-7, 2017}}, \eprint{arXiv:1709.00360}.


\end{thebibliography}

\end{document}